\pdfoutput=1
\documentclass[11pt,twoside,a4paper,cmspaper,final,collab]{cms-tdr}
\def\svnVersion{46a67f0}\def\svnDate{2026/07/10}\def\cmsCernNoTag{CERN-EP-2025-191}\def\cmsCernDate{\today}\def\cmsMessage{Published in the Journal of High Energy Physics as \href{http://dx.doi.org/10.1007/JHEP07(2026)008}{\doi{10.1007/JHEP07(2026)008}.}}
\begin{document}\cmsNoteHeader{EXO-22-006}

\newlength\cmsFigWidth
\ifthenelse{\boolean{cms@external}}{\setlength\cmsFigWidth{0.49\textwidth}}{\setlength\cmsFigWidth{0.65\textwidth}}

\newcommand{\dbse}{\ensuremath{\delta_{\PQb\PQs}}\xspace}
\newcommand{\mll}{\ensuremath{m_{\Pell\Pell}}\xspace}
\newcommand{\METmm}{\ensuremath{\ptmiss/m_{\PGm\PGm}}\xspace}
\newcommand{\METll}{\ensuremath{\ptmiss/\mll}\xspace}
\newcommand{\SR}{\ensuremath{\mathrm{SR_{\PQb}^{\PGm\PGm}}}\xspace}
\newcommand{\CRmmj}{\ensuremath{\mathrm{CR_{\text{j}}^{\PGm\PGm}}}\xspace} 
\newcommand{\CReeb}{\ensuremath{\mathrm{CR_{\PQb}^{\Pe\Pe}}}\xspace} 
\newcommand{\CReej}{\ensuremath{\mathrm{CR_{\text{j}}^{\Pe\Pe}}}\xspace} 
\newcommand{\SRTwo}{\ensuremath{\mathrm{SR_{\PQb+\text{j}/\PQb}^{\PGm\PGm}}}\xspace} 
\newcommand{\CRmmjTwo}{\ensuremath{\mathrm{CR}_{\text{2j}}^{\PGm\PGm}}\xspace} 
\newcommand{\CReebTwo}{\ensuremath{\mathrm{CR}_{\PQb+\text{j}/\PQb}^{\Pe\Pe}}\xspace} 
\newcommand{\CReejTwo}{\ensuremath{\mathrm{CR}_{\mathrm{2j}}^{\Pe\Pe}}\xspace} 
\newcommand{\HTLT}{\ensuremath{H_{\mathrm{T}}-L_{\mathrm{T}}}\xspace}
\newcommand{\LT}{\ensuremath{L_{\mathrm{T}}}\xspace}
\newcommand{\RelMET}{\METmm}

\cmsNoteHeader{EXO-22-006}
\title{Search for a new neutral gauge boson produced in association with one or two b jets and decaying into a pair of muons in proton-proton collisions at \texorpdfstring{$\sqrt{s}=13\TeV$}{sqrt(s) = 13 TeV}}

\date{\today}

\abstract{
 A search for a new neutral gauge boson, \PZpr, produced in association with one or two jets, including at least one \PQb jet, and decaying into a pair of muons is presented. The analysis uses proton-proton collision data collected with the CMS detector at $\sqrt{s} = 13\TeV$, corresponding to an integrated luminosity of 138\fbinv. No significant deviation from background expectations is observed.
 Upper limits at 95\% confidence level on the product of cross section, branching fraction to dimuons, acceptance, and efficiency, from 0.2 to 2\unit{fb}, are set for \PZpr boson masses between 125 and 350\GeV.
 Process-dependent products of acceptance and efficiency, and model-independent limits on the signal yield are provided.
 These are the only results to date in the 125--200\GeV mass range and the most stringent for \PQb quark fusion production modes in the 200--350\GeV range, complementing inclusive \PZpr boson searches.}

\hypersetup{
pdfauthor={CMS Collaboration},
pdftitle={Search for a new neutral gauge boson produced in association with one or two b jets and decaying into a pair of muons in proton-proton collisions at sqrt s = 13 TeV},
pdfsubject={CMS},
pdfkeywords={CMS, BQF, LFUV, Z prime, EXO, lepton flavor universality violation}
}

\maketitle 

\clearpage
\section{Introduction}

The standard model (SM) of particle physics describes a wide range of observed phenomena. However, it is not an ultimate theory of nature as, for example, the SM fails to explain neutrino masses, the matter-antimatter asymmetry of the universe, and the nature of dark matter \cite{PhysRevD.110.030001}. 
In one of the simplest extensions, an additional $\mathcal{U}(1)$ symmetry group gives rise to a new neutral gauge boson, \PZpr, analogous to the SM \PZ boson.  A \PZpr boson is proposed in various models of physics beyond the SM such as $\mathcal{U}(1)_\mathrm{X}$ extensions~\cite{PhysRevD.105.115030, Allanach:2024jls}, Randall--Sundrum models \cite{PhysRevLett.83.3370, PhysRevLett.83.4690, PhysRevD.63.075004}, and grand unified theories \cite{PhysRevD.33.2010, PhysRevD.34.1530}. Such \PZpr bosons could be produced in proton-proton (pp) collisions at the CERN LHC.

In recent years, potential effects beyond the SM have been examined through precision measurements of \PB meson decay rates \cite{PhysRevLett.109.101802,PhysRevD.88.072012, LHCb:2015gmp, PhysRevD.92.072014, PhysRevD.94.072007, PhysRevLett.118.211801, PhysRevD.97.012004, PhysRevLett.120.171802, PhysRevD.97.072013,PhysRevD.108.032002,Aaij_2023}. These effects can be roughly categorized into neutral-current anomalies $R(\PKst)$ and charged-current anomalies $ R(\PDst)$, $R$ being a branching fraction ratio between decay modes with different-flavored lepton daughters. These ratios measure the violation of lepton flavor universality in $\PQb \to \PQs \ell \ell$ and $\PQb \to \PQc \ell \nu_\ell$ transitions, respectively. These measurements have led to proposals of various theoretical models of heavy neutral and charged gauge bosons (\PZpr and \PWpr, respectively) with nonuniversal fermion couplings. Anomalies in $R(\PKst)$ motivate the search for a \PZpr boson that preferentially couples to third-generation fermions \cite{PhysRevD.97.075035,Allanach:2019mfl}. Updated measurements from the LHCb experiment \cite{Aaij_2023, PhysRevD.108.032002} have removed tension originally observed in tests of lepton flavor universality in $\PQb \to \PQs \ell \ell$ transitions that motivated this low-mass \PZpr signal model. More recently, however, an additional spontaneously broken $\mathcal{U}(1)_\mathrm{X}$ gauge group with family-dependent charges has been proposed that is consistent with the latest $R(\PKst)$ measurements along with a TeV-scale \PZpr boson \cite{Allanach:2024jls}. This motivates the current search (\ie, using the dimuon channel only).

\begin{figure}[htbp]
    \begin{tabular}{ccc}
        \raisebox{-0.5\height}{\includegraphics[width=.32\textwidth]{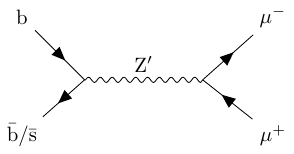}}
        &
        \raisebox{-0.5\height}{\includegraphics[width=.32\textwidth]{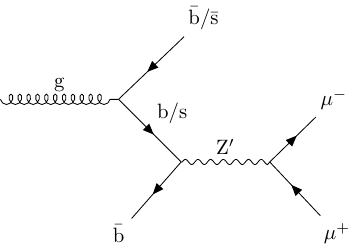}}
        &
        \raisebox{-0.5\height}{\includegraphics[width=.32\textwidth]{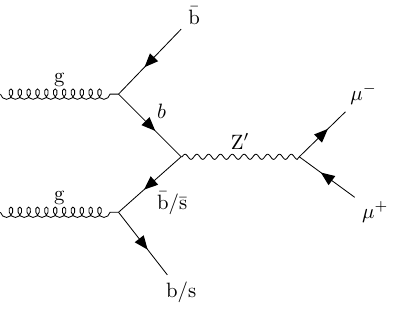}}        
    \end{tabular}
   \caption{Representative Feynman diagrams of \PZpr boson production with bottom-bottom or bottom-strange quark fusion with decays into a dimuon final state. 
   Production at tree level (left), single associated initial-state radiation (ISR) jet production (middle), and two associated ISR jet production (right) are shown.}
   \label{fig:Feynman}
\end{figure}

Searches for a \PZpr boson inclusive in associated jet multiplicity have been performed by both the ATLAS and CMS Collaborations in dielectron and dimuon final states using data collected during the LHC Run 2~\cite{201968, 2020-04-dilepton-final-states,CMS-PAS-EXO-19-019}, which spanned the years 2015--2018.  These searches assume flavor universality of the couplings to all quark and lepton generations, and the leading order (LO) gauge boson production is driven by light-quark initial states. Assuming a \PZpr boson with the same couplings to the third-generation fermions as the SM \PZ boson, bounds on the \PZpr boson mass ($m_{\PZpr}$) are found for the $\PZpr \to \Pgt^{+}\Pgt^{-}$ decay mode, excluding $m_{\PZpr}$ up to 2.42 (ATLAS)~\cite{ATLAS_2018_ditau} and 3.5\TeV (CMS)~\cite{CMS_2025_tau_pairs}$,$ and for the $\PZpr \to \PQb\PAQb$ decay mode, excluding $m_{\PZpr}$ up to 2.7 (ATLAS)~\cite{ATLAS_2020_dijet} and 2.4\TeV (CMS)~\cite{CMS_2023_btagged_dijet}. Models that prefer third-generation quark production modes, such as those in Ref. ~\cite{PhysRevD.97.075035}, which proposes a \PZpr boson decaying to muons that could accommodate flavor anomalies and be detected at the LHC. We probe the decay to muons of a \PZpr boson (with no coupling to electrons) produced via bottom-bottom or bottom-strange quark fusion as a benchmark model. A \PZpr decay into a tau lepton pair is not considered. Consequently, models in Refs.~\cite{Lynch_2001, Chivukula_1994, Muller_1996}, which propose \PZpr bosons decaying to tau leptons, are not directly excluded. Diagrams for the \PZpr boson production and decay are shown in Fig.~\ref{fig:Feynman}.

The search presented here for a \PZpr boson in the mass range of 125--350\GeV complements the search of Ref.~\cite{CMS:2023nzs}, which explored \PZpr boson masses in the range of 350\GeV to 2\TeV. The present study is optimized for a \PZpr boson scenario with exclusive bottom quark couplings in the quark sector and possible flavor-changing neutral current (FCNC) bottom-strange couplings (\dbse). The presence of the \dbse coupling adds diagrams directly accessing the strange quark PDF of the proton for \PZpr boson production, as well as 2-jet diagrams where one of the outgoing jets is a strange quark jet. Increasing \dbse enhances the production cross section and increases the fraction of events with associated light jets.

 In this benchmark scenario, the hypothetical spin-1 narrow \PZpr resonance couples exclusively to muons and muon neutrinos in the lepton sector. We therefore search for a narrow resonance, i.e., with the width less than 1\% of its mass. It is produced in association with one or more jets where at least one is \PQb tagged, and decays into a pair of muons. Dielectron events are treated as a control region (CR).

The search does retain some sensitivity to flavor-universal models. For a sequential-SM~\cite{Langacker:2008yv} \PZpr boson with flavor-universal quark couplings most of the parameter space has already been excluded for masses ranging from 150\GeV to 5.15\TeV \cite{Aad_2014, 201968, CMS-PAS-EXO-19-019}. The sensitivity of the search reported here for flavor-universal lepton couplings is reduced because of the use of dielectron CRs for the background prediction. This limitation will be described in Section~\ref{sec:Results}.

The analysis is performed on LHC Run 2 pp collision data at a center-of-mass energy of 13\TeV collected during 2016--2018, and corresponding to an integrated luminosity of 138\fbinv.
As cross section measurements of Drell--Yan (DY) plus single or double \PQb jet processes have found significant deviations from simulation predictions \cite{PhysRevD.105.092014, ATLAS:2020juj}, the background estimation strategy is fully based on control samples in data. Results are presented in a model-independent manner, covering a range of hypothetical \PZpr boson production with different final states for both pure bottom and FCNC bottom-strange quark couplings. The processes are grouped into seven categories for the interpretation of our results according to the matrix element (ME) calculation of the process. For the left diagram in Fig.~\ref{fig:Feynman}, 0\PQb(\PQb-\PQs) and 0\PQb(\PQb-\PQb) denote \PQb-\PQs and \PQb-\PQb fusion, respectively, with no associated partons. For the middle diagram, 1\PQb denotes a single associated \PQb parton, and 1\PQs denotes a single associated \PQs parton. For the right diagram, 1\PQb\!+1\PQs denotes associated \PQb and \PQs partons, 2\PQb denotes two associated \PQb partons, and 2\PQs denotes two associated \PQs partons arising from a similar diagram where an ISR gluon splits into an \PQs{}\PAQs parton pair.

This paper is structured as follows: Section \ref{sec:CMSdetectorEventReco} describes the CMS detector and the event reconstruction. Section \ref{sec:Samples} introduces the data used and the simulated samples generated. Section \ref{sec:EventSelection} deals with the event selection and categorization. Section \ref{sec:BackgroundEstimation} describes the background estimation, while Section \ref{sec:SysUnc} lists all systematic uncertainties considered. Results are shown in Section \ref{sec:Results}, with a summary provided in Section \ref{sec:Summary}. Tabulated results can be found in the HEPData record for this analysis \cite{HEPData}.

\section{The CMS detector and event reconstruction\label{sec:CMSdetectorEventReco}}

The CMS apparatus~\cite{CMS:2008xjf,CMS:2023gfb} is a multipurpose, nearly hermetic detector, designed to trigger on~\cite{CMS:2020cmk,CMS:2016ngn,CMS:2024aqx} and identify electrons, muons, photons, and (charged and neutral) hadrons~\cite{Sirunyan_2021,CMS:2018rym,CMS:2014pgm}. Its central feature is a superconducting solenoid of 6\unit{m} internal diameter, providing a magnetic field of 3.8\unit{T}. Within the solenoid volume are a silicon pixel and strip tracker, a lead tungstate crystal electromagnetic calorimeter (ECAL), and a brass and scintillator hadron calorimeter (HCAL), each composed of a barrel and two endcap sections. Forward calorimeters extend the pseudorapidity coverage provided by the barrel and endcap detectors. Muons are reconstructed using gas-ionization detectors embedded in the steel flux-return yoke outside the solenoid. More detailed descriptions of the CMS detector, together with a definition of the coordinate system used and the relevant kinematic variables, can be found in Refs.~\cite{CMS:2008xjf,CMS:2023gfb}. 

Events of interest are selected using a two-tiered trigger system. The first level (L1), composed of custom hardware processors, uses information from the calorimeters and muon detectors to select events at a rate of around 100\unit{kHz} within a fixed latency of about 4\mus~\cite{CMS:2020cmk}. The second level, known as the high-level trigger, consists of a farm of processors running a version of the full event reconstruction software optimized for fast processing, and reduces the event rate to around 1\unit{kHz} before data storage~\cite{CMS:2016ngn}. Events are selected using either (i)~a trigger requiring a single muon, with a minimum \pt\ threshold of 50\GeV, updated in 2017 and 2018 to use a slightly different reconstruction method better optimized for high-\pt\ muons, or (ii)~a double-electron trigger with a \pt\ requirement of 33\GeV per electron, which was lowered to 25\GeV during 2017 and 2018. Single-muon triggers are used instead of double-muon triggers because they offer higher acceptance, and the primary advantage of double muon triggers, the lower \pt threshold, is not required.

The global event reconstruction, also called particle-flow (PF) event reconstruction~\cite{CMS:2017yfk}, aims to reconstruct and identify each individual particle in an event, with an optimized combination of all subdetector information. In this process, the identification of the particle type (photon, electron, muon, charged hadron, neutral hadron) plays an important role in the determination of the particle direction and energy. Photons are identified as ECAL energy clusters not linked to the extrapolation of any charged-particle trajectory to the ECAL. Electrons are identified as a primary charged-particle track and potentially many ECAL energy clusters corresponding to this track extrapolation and to possible bremsstrahlung photons. Muons are identified as tracks in the central tracker consistent with either a track or several hits in the muon system, and associated with calorimeter deposits compatible with the muon hypothesis. Charged hadrons are identified as charged-particle tracks neither identified as electrons, nor as muons. Finally, neutral hadrons are identified as HCAL energy clusters not linked to any charged-hadron trajectory, or as a combined ECAL and HCAL energy excess with respect to the expected charged-hadron energy deposit.

Hadronic jets are clustered from reconstructed PF particles using the anti-\kt algorithm~\cite{Cacciari:2008gp, Cacciari:2011ma} with a distance parameter of 0.4. Jet momentum is determined as the vectorial sum of all particle momenta in the jet, and is found from simulation to be, on average, within 5 to 10\% of the true particle-level momentum. Additional pp interactions within the same or nearby bunch crossings (pileup) can contribute additional tracks and calorimetric energy depositions to the jet momentum. To mitigate this effect, charged-particles identified to be originating from pileup vertices are discarded and an offset correction is applied to correct for remaining contributions. Jet energy corrections are derived from simulation to bring the measured response of jets to that of particle-level jets on average, and in situ measurements of the momentum balance in dijets, multijets, and Z+jets events are used to account for any residual differences in the jet energy scale between data and simulation~\cite{CMS:2016lmd}. The jet energy resolution amounts typically to 15--20\% at 30\GeV, 10\% at 100\GeV, and 5\% at 1\TeV~\cite{CMS:2016lmd}. We refer to the group of either type of lepton (\Pe or \PGm) or jet, irrespective of \PQb tagging, as an object, from hereon out. A tight lepton veto is applied to ensure jets are not misreconstructed leptons and a $\Delta R = \sqrt{(\Delta \eta)^2 + (\Delta\phi)^2} > 0.4$ requirement is applied to the nearest electron or muon, where $\Delta\eta$ and $\Delta\phi$ are the differences in the pseudorapidity and azimuthal angle, respectively, between the two objects. Additional selection criteria are applied to each jet to remove jets potentially dominated by anomalous contributions from various subdetector components or reconstruction failures~\cite{CMS-PAS-JME-16-003}. 

The missing transverse momentum vector \ptvecmiss is computed as the negative vector sum of the transverse momenta of all the PF candidates in an event, and its magnitude is denoted as \ptmiss~\cite{CMS:2019ctu}. The \ptvecmiss is modified to account for corrections to the energy scale of the reconstructed jets in the event. 

Jets containing b quarks can be identified utilizing the relatively long b quark lifetime and identifying a secondary vertex within the jet, as well as using jet shape variables~\cite{BTV-16-002}. We utilize the medium working point of the \textsc{DeepJet} algorithm~\cite{Bols:2020bkb, CMS-DP-2023-005} defined by a 1\% misidentification probability for jets originated from light-quarks or gluons, yielding an efficiency of ${\approx}90\%$ to identify jets originated from \PQb quarks as \PQb-tagged. All jets are required to fulfill $\abs{\eta}<2.4$, a tight jet identification selection to suppress detector noise, and for jet $\pt < 50\GeV$, a boosted decision tree discriminator is used to reject pileup jets if the score passes a certain threshold (tight working point) \cite{CMS:2020ebo, CMS-PAS-JME-16-003}. The \PQb-tagged jets are subject to a requirement of $\pt > 20\GeV$, while non-\PQb-tagged jets must satisfy $\pt > 30\GeV$. The lower \pt requirement for \PQb jets allows for greater acceptance of the soft associated \PQb jet production and is possible since the \PQb-tagged jet selection retains purity at lower \pt. Lastly, all jets must be separated from any identified isolated electron or muon by at least $\Delta R>0.4$. Scale factors to correct for the differences between data and simulation are applied to pileup and \PQb tagging criteria~\cite{BTV-16-002}.

The electron momentum is estimated by combining the energy measurement in the ECAL with the momentum measurement in the tracker. The relative energy resolution for electrons with$\pt \approx 45\GeV$ from $\PZ \to \Pe \Pe$ decays ranges from 1.4 to 3\%. It is generally better in the barrel region than in the endcaps, depends on the bremsstrahlung energy emitted by the electron as it traverses the material in front of the ECAL, and the data-taking year~\cite{Sirunyan_2021,CMS-DP-2020-021}. This results in a dielectron mass resolution for $\PZ \to \Pe \Pe$ decays in the ranges  1.2--2.0\% or 2.2--3.2\% when both electrons are in the ECAL barrel or in the ECAL endcaps, respectively

Muons are measured in the pseudorapidity range $\abs{\eta} < 2.4$. The single-muon trigger efficiency exceeds 90\% over the full $\eta$ range, and the efficiency to reconstruct and identify muons is greater than 96\%. Matching muons to tracks measured in the silicon tracker results in a relative transverse momentum resolution, for muons with \pt up to 100\GeV, of 1\% in the barrel and 3\% in the endcaps. The \pt resolution in the barrel is better than 7\% for muons with \pt up to 1\TeV~\cite{CMS:2018rym}.

Electrons and muons in this analysis are required to fulfill $\pt > 53\GeV$ and $\abs{\eta}<2.4$, slightly above the 50\GeV muon trigger threshold. We require tight quality criteria with misidentification probabilities of 1\% for electrons, using high-energy electron pair (HEEP) identification~\cite{Sirunyan_2021}, and $0.1\%$ for muons, using the cutoff-based high-\pt identification~\cite{CMS:2018rym}. Moreover, both types of leptons are required to be isolated, where the relative isolation is measured by the \pt sum of all additional PF candidates in a cone around the lepton relative to the lepton \pt~\cite{Khachatryan:2015hwa, CMS:2018rym}. The cone size is $\Delta R<0.3$ for electrons and $<0.4$ for muons. The isolation is part of a multivariate analysis (MVA) discriminant for electrons with the aforementioned misidentification rate and a total efficiency of reconstruction, identification, and the isolation requirement of 85--90\%~\cite{Sirunyan_2021}. For muons, relative isolation is evaluated independently of the identification working point and is required to be less than $0.1$. Lepton isolation is corrected for neutral-hadron contributions from pileup \cite{CMS:2018rym, Sirunyan_2021}. This selection is synchronized with the CMS dilepton final-state search \cite{CMS-PAS-EXO-19-019} and utilizes the high-\pt identification developed for that analysis. Muon and electron momentum scale factors are applied to amend differences in momentum reconstruction between data and simulation. Both lepton types are corrected in simulation with scale factors for differences in identification, isolation, and reconstruction efficiencies to data.

An additional loose selection of electrons or muons is used to veto events with extra leptons. A loose set of quality criteria is applied, using the MVA identification for electrons~\cite{Sirunyan_2021}, as the HEEP identification~\cite{Sirunyan_2021} does not allow for selecting sufficiently low-\pt electrons, and the cutoff-based high-\pt identification for the muons along with the requirements  $\pt > 10\GeV$ and $\abs{\eta}<2.4$\cite{CMS:2018rym}.

Table \ref{tab:objSelection} summarizes these selection requirements.

\begin{table}
    \centering
	\caption{Summary of object selection requirements on leptons, veto leptons, and jets passing and failing \PQb jet identification. All objects have a maximum $\abs{\eta}$ of 2.4. The efficiencies of the respective identification working points are detailed in the text of Section \ref{sec:CMSdetectorEventReco}.}
       \begin{tabular}{ccc}
            \hline
            Object & Identification & Minimum \pt [\GeVns{}]\\
            \hline
            \PGm  & Tight          & 53 \\
            \Pe    & Tight          & 53 \\[1pt]
            Veto \PGm & Loose      & 10 \\
            Veto e & Loose        & 10 \\[1pt]
            \PQb jet  & Pass medium \PQb tag   & 20 \\
            Non-\PQb jet & Fail medium \PQb tag & 30 \\
            \hline
       \end{tabular}
   
       \label{tab:objSelection}
   \end{table}

\section{Data and simulated samples\label{sec:Samples}}

Signal Monte Carlo (MC) samples were prepared with \MGvATNLO~\cite{MadGraph}~version 2.4.2 (2.6.0) for 2016 (2017 and 2018) at leading order (LO) with the NNPDF3.1 \cite{NNPDF:2017mvq} parton distribution functions (PDFs) in the five flavor scheme. The MLM technique \cite{MLM_2007} is used to eliminate the double counting between jets from the matrix element calculations and parton shower. The signal process is generated with up to two additional partons. The Universal FeynRules Output (UFO) ~\cite{Christensen_2009} configuration file for the signal process was privately prepared and can be found in the HEPData record for this analysis \cite{HEPData}. Signals are generated at \PZpr boson masses of 125, 150, 175, 200, 250, 300, and 350\GeV, and with \dbse values of 0.04, 0.5, and 1.0, with resonance widths of less than 1\% of the \PZpr boson mass. This is lower than the reconstructed dimuon mass resolution, which ranges from 1.7\GeV at 125 GeV to 7.2 GeV at 350\GeV.  Different  \dbse  samples change the proportion of different production categories, each with its unique product of acceptance and efficiency, but not the dimuon mass shape. The \dbse values were chosen to cover different production categories \cite{PhysRevD.97.075035}.

DY + jets processes are simulated at next-to-leading order (NLO) using \MGvATNLO version 2.4.2 (2.6.5) for 2016 (2017 and 2018) data taking, with jets from the matrix element calculations and from parton shower matched using the FxFx scheme \cite{FxFx}. Single top quark, top quark pair production ($\PQt\PQt$), and diboson  $(\PV\PV=\PW\PW, \PW\PZ, \PZ\PZ)$ samples are simulated at NLO using the \POWHEG v2.0 generator. Additional NLO DY samples generated with \POWHEG are used to increase the event count of the \MGvATNLO~samples in the low-mass regions where the selections are most stringent. In the \POWHEG samples, the jet matching is handled internally.

The NNPDF3.0 \cite{NNPDF:2014otw} parton distribution functions (PDFs) are used for 2016, while NNPDF3.1 \cite{NNPDF:2017mvq} is used for the 2017 and 2018 data taking. The matrix element generators are interfaced with \PYTHIA 8.226 (8.230) \cite{pythia8} for parton showering and underlying event description using the CUETP8M1 tune~\cite{CMS:2015wcf} (CP5 tune~\cite{Sirunyan:2019dfx}) for 2016 (2017 and 2018) samples.

Pileup is included in the simulation, as generated with \PYTHIA, with its multiplicity distribution adjusted to match that observed in data. 
All simulated events are processed by a \GEANTfour-based \cite{AGOSTINELLI2003250}~simulation of the CMS detector, in the same version of the CMS software as used in data reconstruction.

\section{Event selection and categorization\label{sec:EventSelection}}

After selecting events that pass the single muon or double-electron triggers, we retain only events with exactly two opposite-sign same-flavor leptons with selection criteria that ensure high purity and veto any event with any further loosely selected leptons of any flavor, as detailed in the object selections described in Section \ref{sec:CMSdetectorEventReco}. To search for a \PZpr boson that decays to muons, with associated \PQb jets, we create dimuon signal regions (SRs) containing at least one b-tagged jet and CRs enhanced with either events containing either electrons, or light-quark, or gluon, jets.

We select events with exactly one or two jets. 
These two jet multiplicity categories are handled separately to distinguish the two types of leading diagrams in their respective jet multiplicity categories for the targeted signal:
a quark-gluon initiated diagram with one final-state \PQb quark in the 1-jet category, and a gluon-gluon initiated diagram with an $\PQs\PQb$ or $\PQb\PQb$ quark pair in the final state for the 2-jet category. The presence of a flavor-violating \dbse coupling adds diagrams directly accessing the strange quark PDF of the proton for \PZpr boson production, as well as 2-jet diagrams where one of the outgoing jets is a strange quark jet. Therefore, the strength of such a coupling changes both the branching fraction between the 1- and 2-jet categories, and their acceptances. Higher jet multiplicities are not included due to of minimal signal acceptance and larger additional background. The SRs and the dielectron \PQb jet enriched CRs require at least one jet to be \PQb tagged, while the light-jet regions require no \PQb-tagged jets. In the labeling of SRs and CRs, dimuon and dielectron regions are denoted with a $\PGm\PGm$ or $\Pe\Pe$ superscript, respectively, and a ``j", ``\PQb{}", ``2j" or ``$\PQb+\mathrm{j}/\PQb$" subscript indicates that the region is a single/double light-quark jet or \PQb jet region. Table \ref{tab:sr_cr} lists the resulting SRs and CRs.

\begin{figure}[h!t]
  \centering
  
  \includegraphics[width=0.49\textwidth]{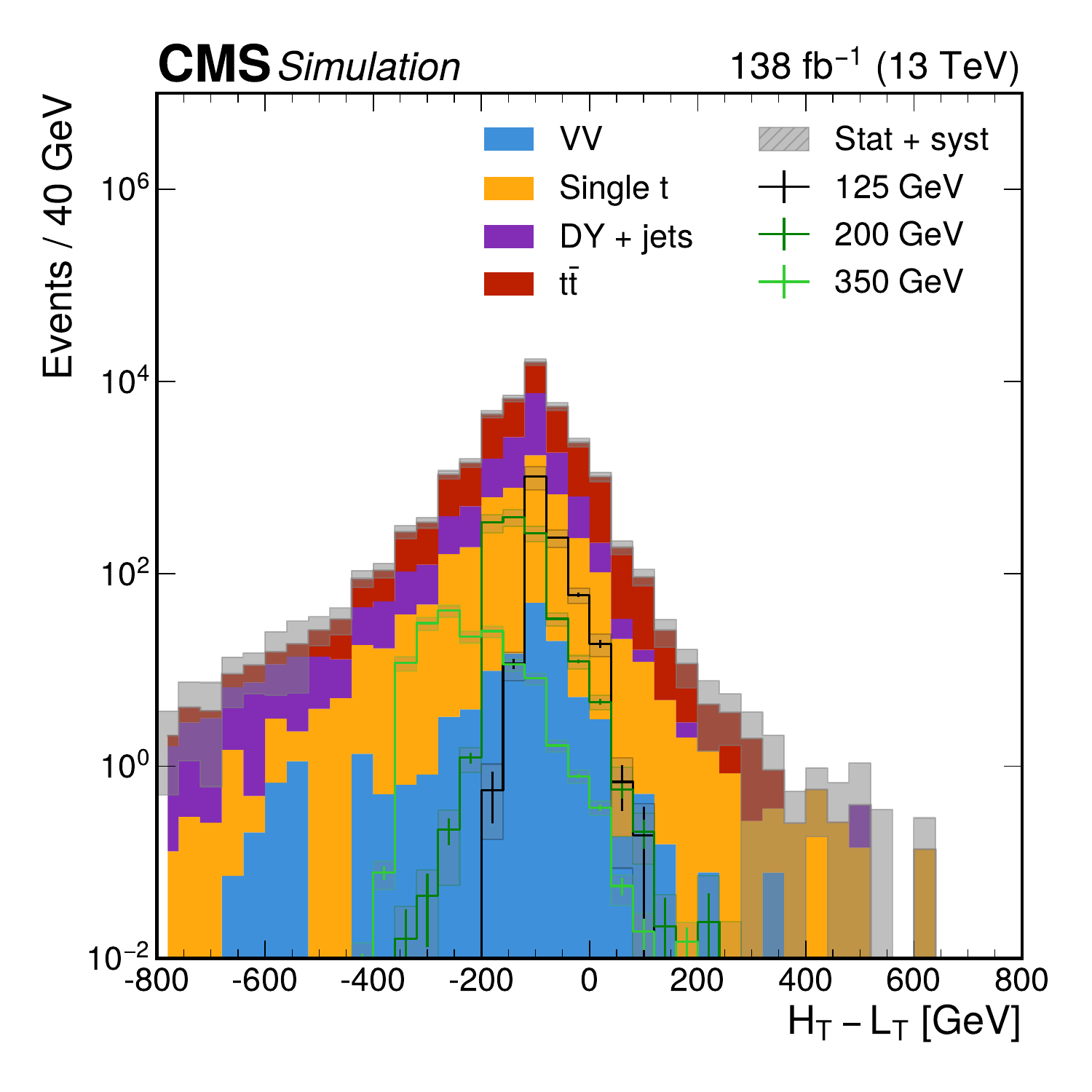}
  \hfill
  \includegraphics[width=0.49\textwidth]{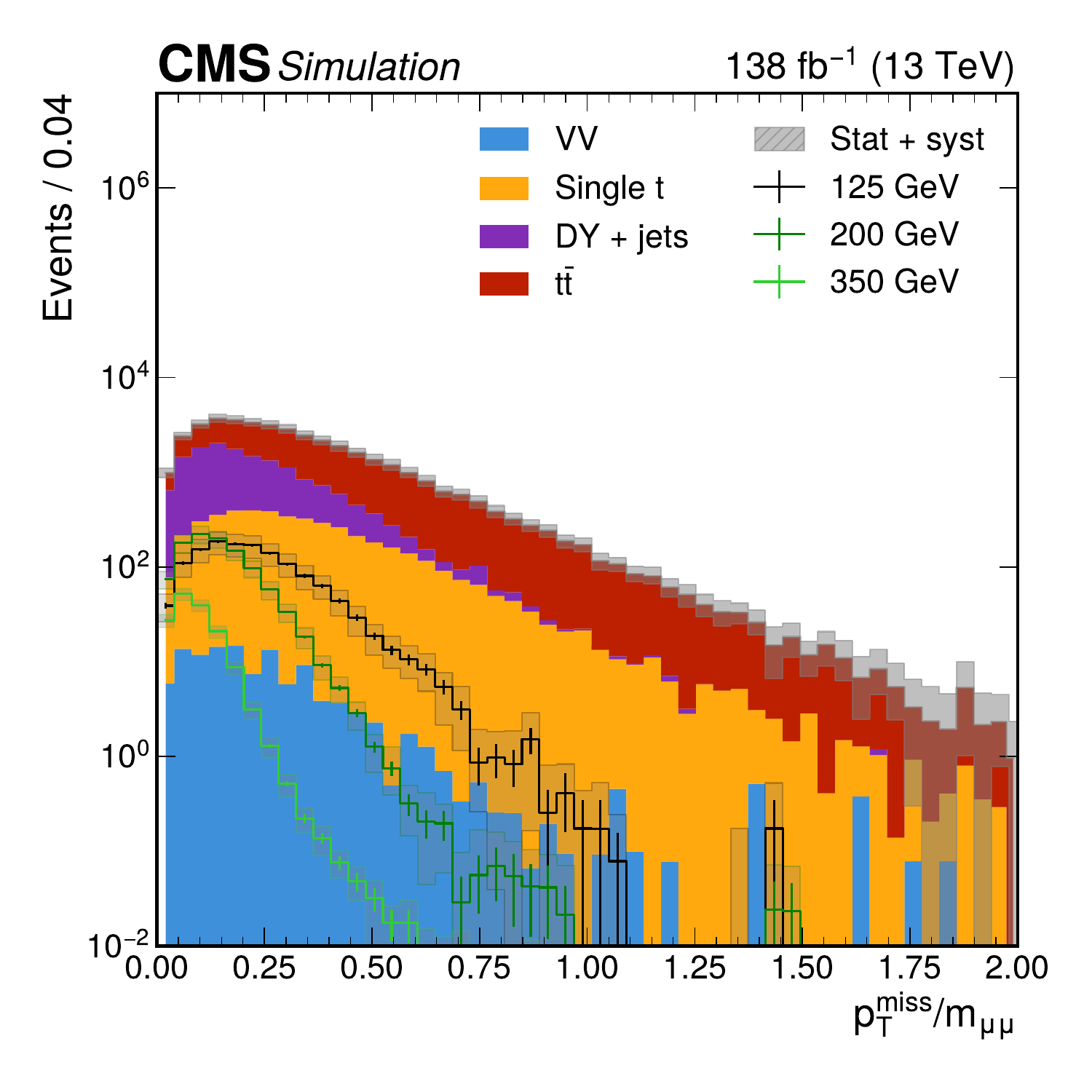}

  \vspace{1mm} 
  
  \includegraphics[width=0.49\textwidth]{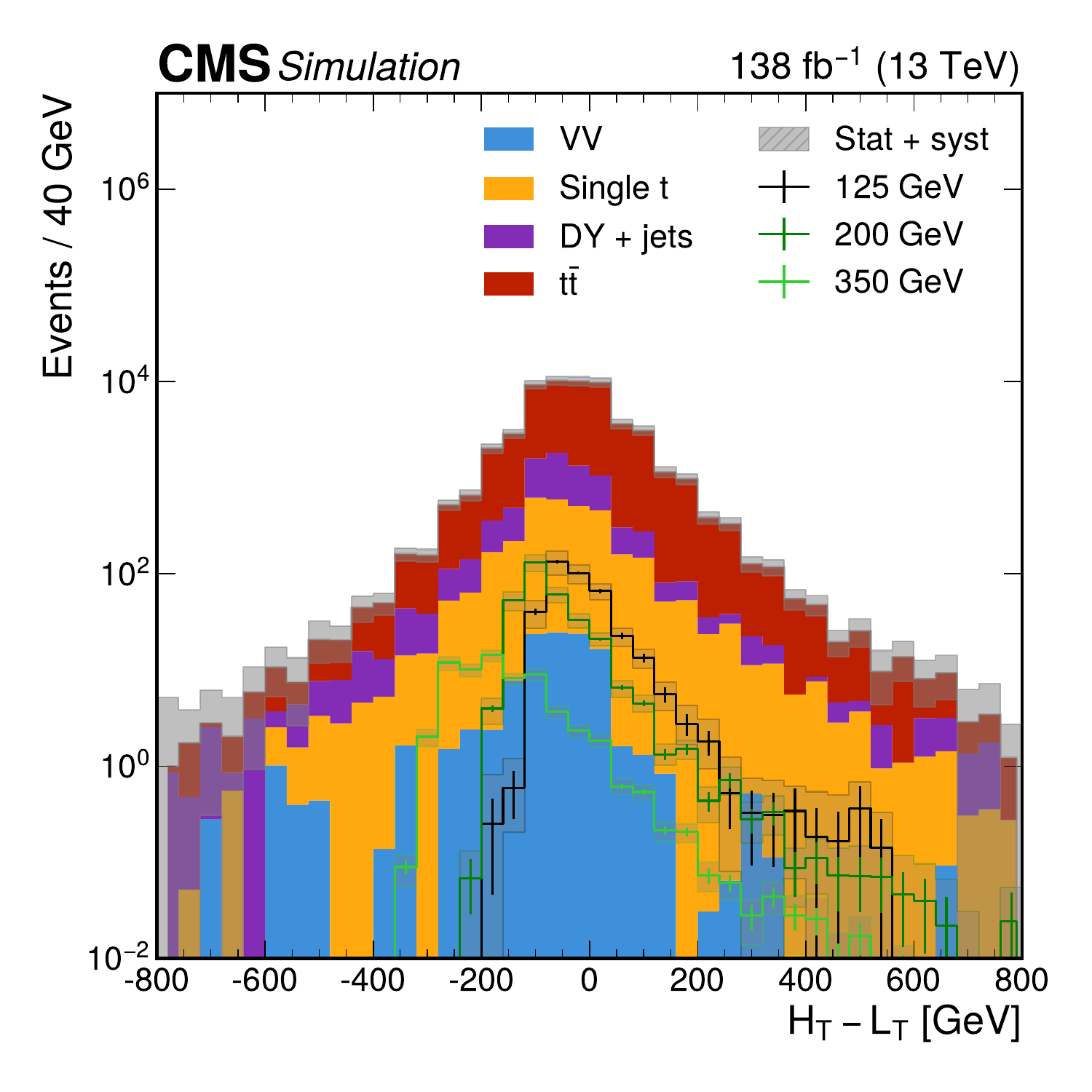}
  \hfill
  \includegraphics[width=0.49\textwidth]{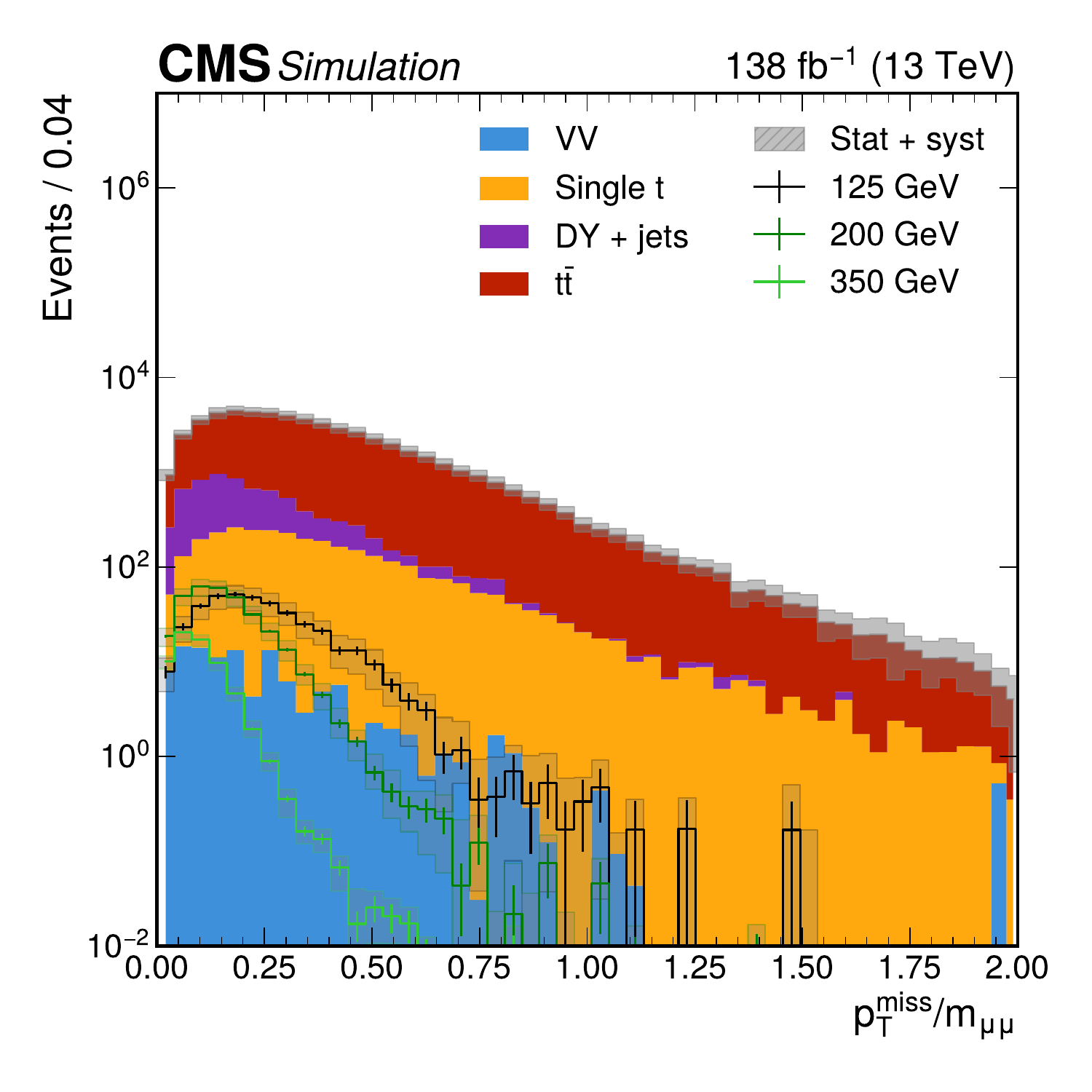}

  \caption{Distributions of the search-specific variables from simulated backgrounds and signals for \PZpr boson masses of 125, 200, and 350\GeV after object selections including a single-muon requirement and categorization into both jet multiplicities. The left column shows the \HTLT~ distributions, and the right column shows the \METll~ distributions. The upper row shows the 1-jet category distributions, and the lower row shows the 2-jet category distributions. Overflow events are not shown. \label{fig:BQFvars}}  
\end{figure}

Further requirements to suppress backgrounds beyond the selections introduced in Section \ref{sec:CMSdetectorEventReco} depend on the dilepton invariant mass \mll~ and the jet multiplicity. Consequently, two-dimensional Bottom Quark Fusion (BQF) variable selections are applied to optimize the analysis as a function of the reconstructed dilepton mass. These are upper bounds on \RelMET\ and on the difference between the scalar sum of jet transverse momenta (\HT) and lepton transverse momenta (\LT), denoted as \HTLT. Distributions of these BQF variables, proposed in Ref.~\cite{PhysRevD.97.075035}, for backgrounds and signals in \SR\ and \SRTwo\ are shown in Fig. \ref{fig:BQFvars}. The signal distributions for the \HTLT\ and \RelMET\ variables shift as a function of the \PZpr boson mass.

We optimize the selections on each BQF variable to maximize the expected significance, defined as the signal counts divided by the square root of the sum of signal and expected background counts, for a range of signal mass hypotheses between 125 and 350\GeV. The reconstructed \PZpr boson mass is used as a proxy of the generated \PZpr boson mass to create two-dimensional selections tuned for optimal performance as a function of the mass. The results of the optimization are shown in Table \ref{tab:evSelection}.

At the lowest dimuon masses in our search range, the background suppression is less effective, as both \HTLT\ and \RelMET\ better separate signal from background for high-\pt\ dimuons. Additionally, the decay products of a low-mass \PZpr boson are generally more similar to those of SM processes, reducing the efficacy of the BQF selections. Conversely, with increasing mass, the background rejection power of the BQF variables improves.

Compared to the variables used for masses above 350\GeV in the high-mass analysis \cite{CMS:2023nzs}, the BQF variables offer less stringent background rejection but higher acceptance. For dilepton masses between 125 and 350\GeV, background events are reduced to 78--42\% of their pre-BQF-variable selection values in \SR\ and to 43--21\% in \SRTwo. Signal events are reduced to 95--84\% in \SR\ and 76--67\% in \SRTwo\ relative to their pre-BQF-variable selection values.

We analyze the dilepton mass range 122--391\GeV so as to cover a \PZpr boson mass range of 125--350\GeV, by expanding the mass window by 3 times the dimuon mass resolution in each direction.

\begin{table*}[htbp]
   \centering
	\caption{Number of jet requirements for the SRs and CRs based on opposite-sign dilepton pair flavor, number of \PQb-tagged jets ($N_{\PQb}$), and the overall number of jets ($N^{\textrm{all}}_{\textrm{jets}}$). Events with more than two leptons passing veto lepton selections of any flavor are discarded. In the chosen benchmark model, the \PZpr boson does not decay into electrons, hence the use of the dielectron CRs. }
  \renewcommand{\arraystretch}{1.0} 

   \begin{tabular}{llcc}
   \hline\\[-7pt]
   $N_{\textrm{b}}$ & $N^{\textrm{all}}_{\textrm{jets}}$ & \MM & \EE\\[4pt]
   \hline\\[-7pt]
   $\geq$1 & $2$ & \SRTwo & \CReebTwo \\[4pt]
   $0$ & $2$ & \CRmmjTwo & \CReejTwo \\[4pt]
   $1$ & $1$ & \SR & \CReeb \\[4pt]
   $0$ & $1$ & \CRmmj & \CReej \\[4pt]
   \hline
  \end{tabular}
  
  \label{tab:sr_cr}
\end{table*}

\begin{table}[ht]
    \centering
    \caption{Summary of additional \mll-dependent event selection requirements on \ptmiss\ and \HTLT~ BQF variables to suppress backgrounds. The jet multiplicity column indicates the maximum value the BQF variables can have to pass the selection for the respective number of jets.}
    \begin{tabular}{ccc}
     \hline
     \multirow{2}*{BQF variable} & \multicolumn{2}{c}{Jet multiplicity}\\
     & 1 & 2\\\hline
    \rule{0pt}{3ex}  \ptmiss$/\mll$ $<$   & $223(\mll/\GeVns)^{-0.23}$  & $67  (\mll/\GeVns)^{-0.04}$  \\
     \HTLT\ $<$ &  $-0.43 \mll+47$\GeV &  $-0.56 \mll+124$\GeV\\
     \hline
    \end{tabular}
    \label{tab:evSelection}
\end{table}

\section{Background estimation\label{sec:BackgroundEstimation}}

In this search, we use an ABCD (matrix) background estimation technique based on parametric fits to the CRs to smooth statistical fluctuations. All previous selections, including the BQF ones, are applied before performing the fits. Jet and electron selections are used to construct the CRs, which are then used to predict the backgrounds in the muon-enriched and \PQb jet enriched SRs.

\begin{figure}[hbt!]
  \centering
  \includegraphics[width=.5\linewidth]{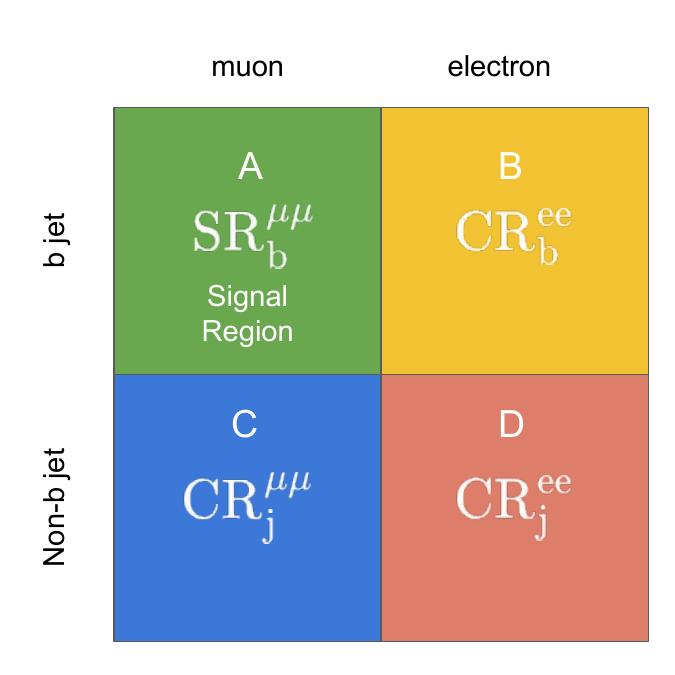}
  \caption{A visualization of the ABCD SRs and CRs. For example, \SR represents the dimuon b jet enriched SRs (\SR\ or \SRTwo) and \CReeb represents the dielectron b jet enriched CRs (\CReeb\ or \CReebTwo).}
  \label{fig:ABCD}
\end{figure}

Figure \ref{fig:ABCD} shows a visualization of the ABCD method. Here, the symbols A, B, C, and D refer to four regions constructed from two uncorrelated selection based on leptons and jets. One of the regions (A) is dominated by signal while the other three CRs (B, C, D) are background-dominated. The background yield in the SR is estimated from the yields in the three CRs with the expression: $N_A = N_B  N_C/N_D$, where $N_X$ is the number of events in region X.

For this analysis, as the benchmark model predicts no decays to electrons, the background shape is extracted from a dielectron CR with at least one b-tagged jet for both the one (\CReeb) and two (\CReebTwo) jet regions, as also indicated in Table \ref{tab:sr_cr}. To correct for differences in the dilepton mass shape and normalization arising from acceptance, resolution, and other differences between electrons and muons, a multiplicative transfer factor per bin in the dilepton mass distribution is determined from functional fits, described below, in non-b-tagged regions: \CRmmj/\CReej~ and \CRmmjTwo/\CReejTwo~ for the one- and two-jet regions, respectively. These per-bin transfer factors are then multiplied per bin by fit values in the dielectron b-tagged regions \CReeb and \CReebTwo to create the predictions for backgrounds in \SR and \SRTwo. Thus, the ABCD background prediction equations are given by:

\begin{equation}
	SR_{\PQb}^{\PGm\PGm} = CR_{\PQb}^{\Pe\Pe}\frac{CR_{\textrm{j}}^{\PGm\PGm}}{CR_{\textrm{j}}^{\Pe\Pe}}, \quad SR_{\PQb+\textrm{j}/\PQb}^{\PGm\PGm} = CR_{\PQb+\textrm{j}/\PQb}^{\Pe\Pe}\frac{CR_{\textrm{2j}}^{\PGm\PGm}}{CR_{\textrm{2j}}^{\Pe\Pe}}.
    \label{eqn:sr_abcd_eqn}
\end{equation}

This method assumes that there is no signal in the CRs. In the benchmark model, signal contamination in the dimuon non-b-tagged regions has been tested and found to be negligible even with a signal ten times the expected exclusion cross section, because of the far higher background rates and lower signal acceptance, even for production with additional associated strange jets. Furthermore, any initial/final-state radiation or PDF effects cancel in the ratio of the regions of different lepton flavors by construction. In addition, a signal injection test into the dielecton b jet enriched CRs was conducted to demonstrate robustness against signal contamination for \PZpr scenarios with electron couplings (as discussed in Section \ref{sec:Results}).

The main backgrounds depend on the associated jet multiplicity and dilepton mass. 
In general, \ttbar is the most relevant background at higher dilepton masses, while in the lower range, DY processes are dominant. 
A subdominant contribution from single top quark production is always present at either multiplicity, and rarer backgrounds, such as $\PV\PV$, contribute at the sub-percent level.
 Most of the DY background is in the 1-jet regions (${\approx}50\%$ DY, ${\approx}40\%$ \ttbar, ${\approx}10\%$ single top quark), while \ttbar~background aggregates preferably in the 2-jet regions (${\approx}20\%$ DY, ${\approx}75\%$ \ttbar, ${\approx}5\%$ single top quark).

All backgrounds are estimated directly from data using the ABCD method rather than from simulations. This is motivated because (i) in cross section measurements of DY production with one or two associated \PQb jets, discrepancies between simulations and data have been observed \cite{PhysRevD.105.092014}, and (ii) uncertainties in the bottom and strange quark PDF contributions would affect predictions for the relative rate of background processes, and thereby the background acceptance in regions with one to two associated \PQb jets.

\begin{figure}[hbt]
\centering
      \includegraphics[width=0.49\textwidth]{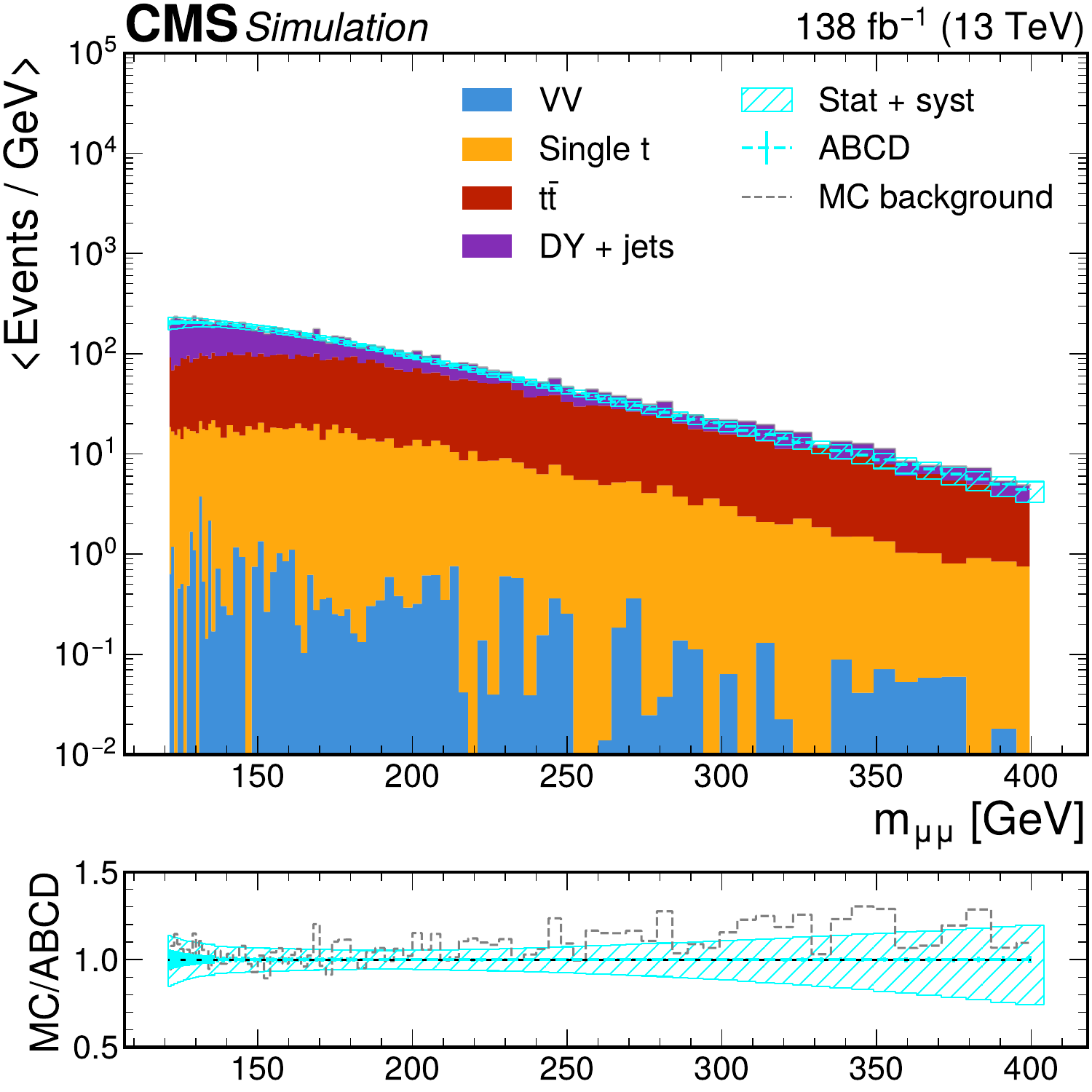}
      \hfill
      \includegraphics[width=0.49\textwidth]{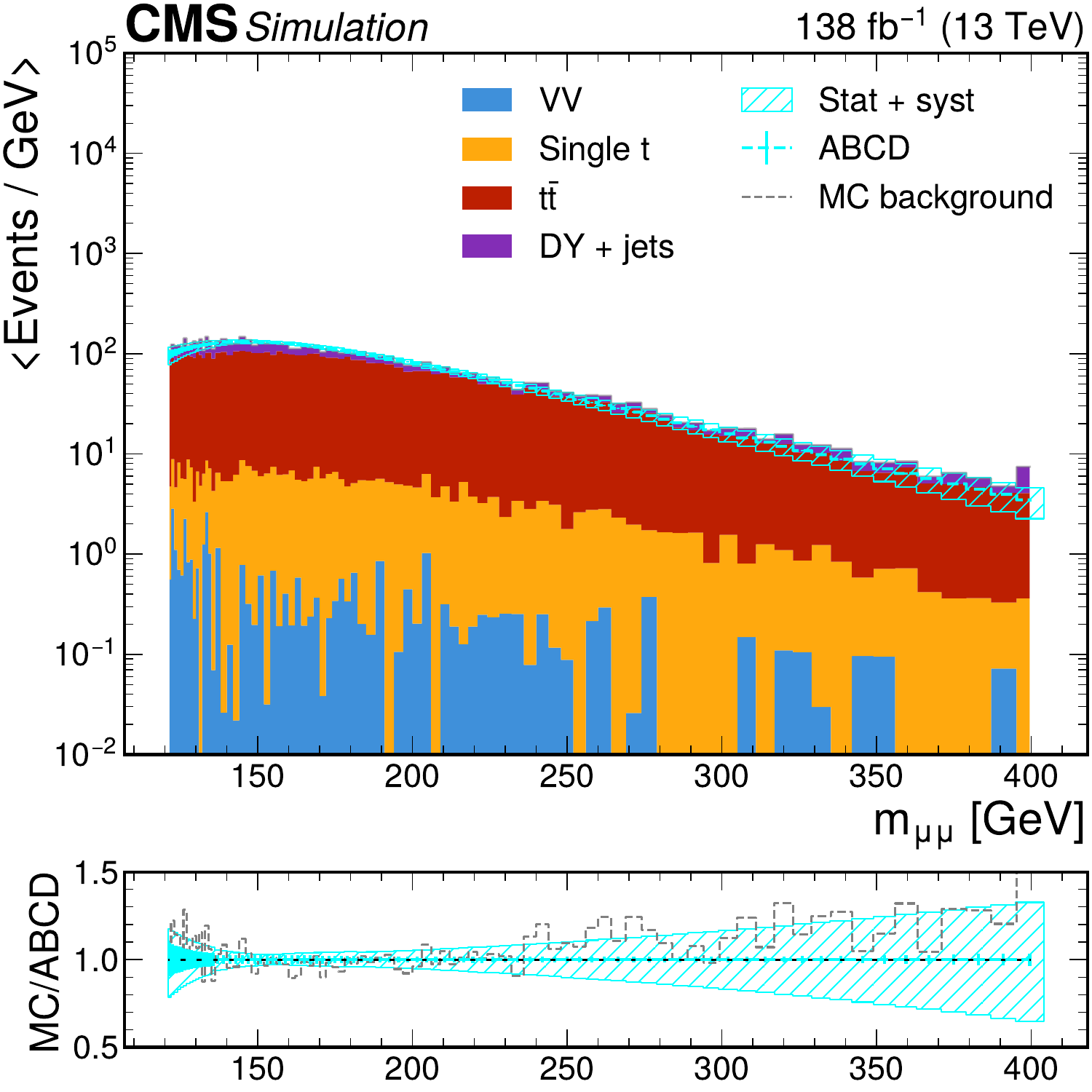}
 \caption{Comparison between the predictions of the \SR~ background from simulation (gray) and the ABCD method (teal) from MC simulation. Events are divided by the bin width. The statistical and systematic uncertainties using the ABCD method are shown around it in a teal hashed area. The ratio of the nominal MC background values (dashed line) to the ABCD prediction is shown as a dashed line in the ratio plot. The MC background uncertainties are not shown for visual clarity.  Left: \SR. Right: \SRTwo. \label{fig:BgrClosure}} 
\end{figure}

\begin{figure}[hbt]
  \centering

  \includegraphics[width=0.48\textwidth]{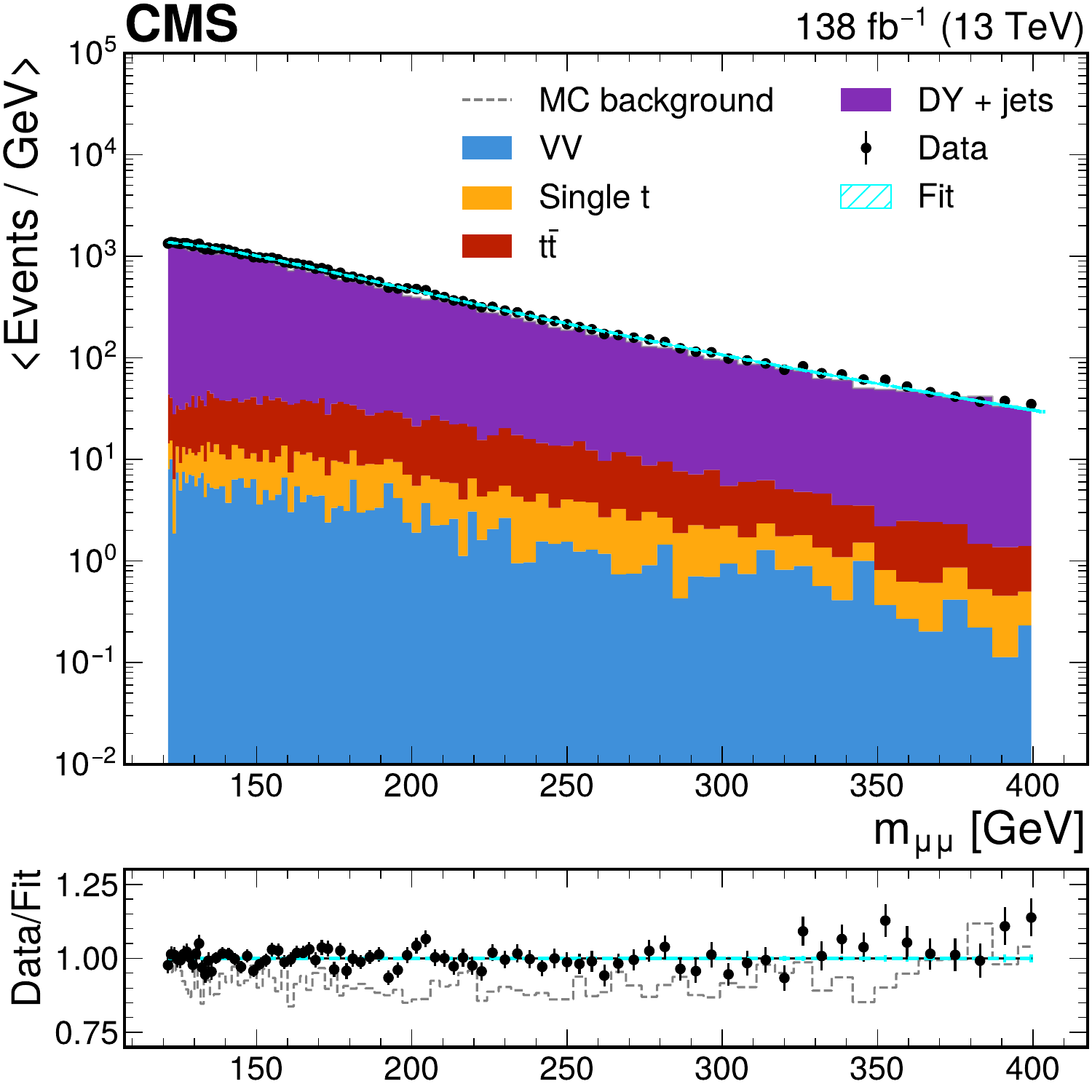}
  \hfill
  \includegraphics[width=0.48\textwidth]{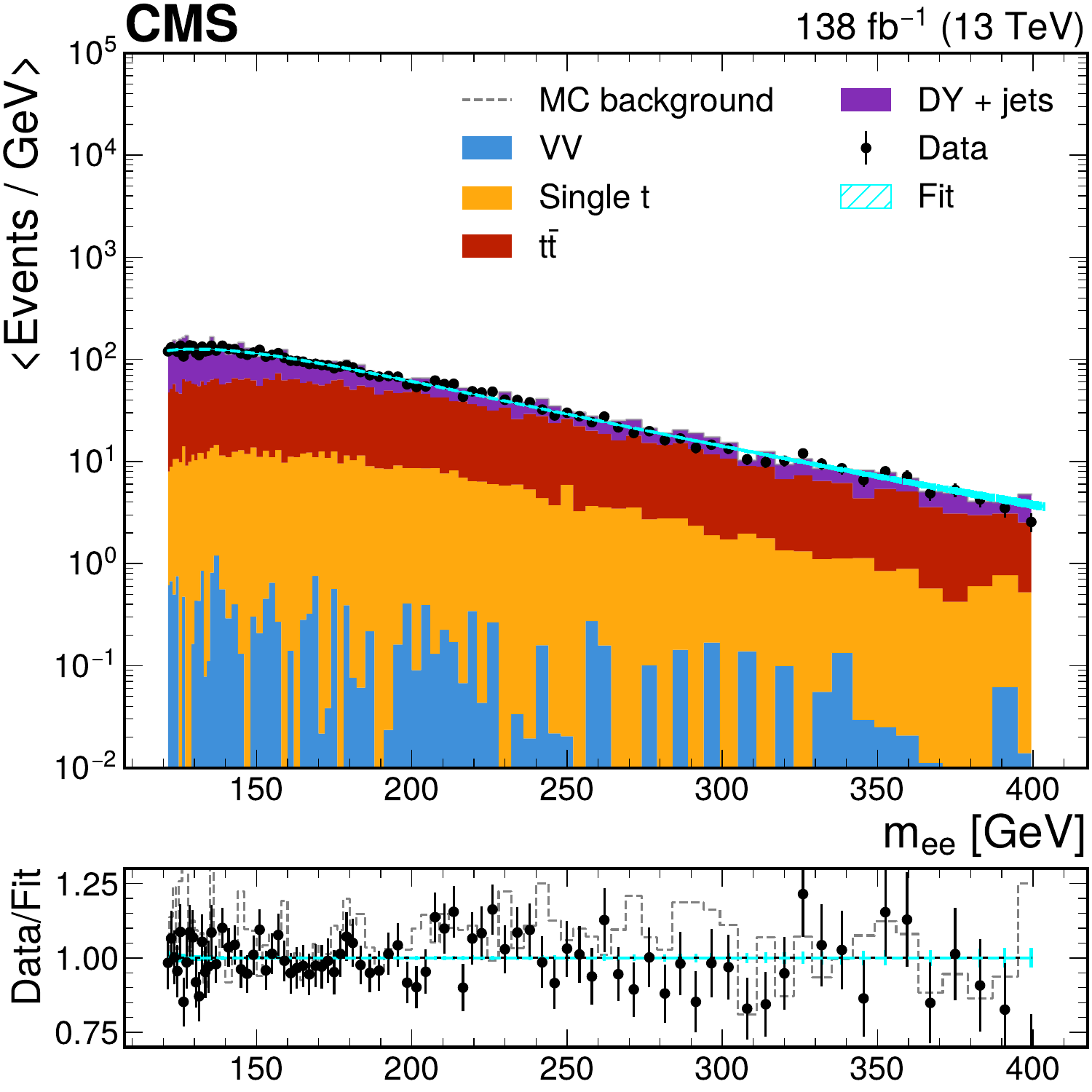}

  \vspace{1mm} 

  \includegraphics[width=0.48\textwidth]{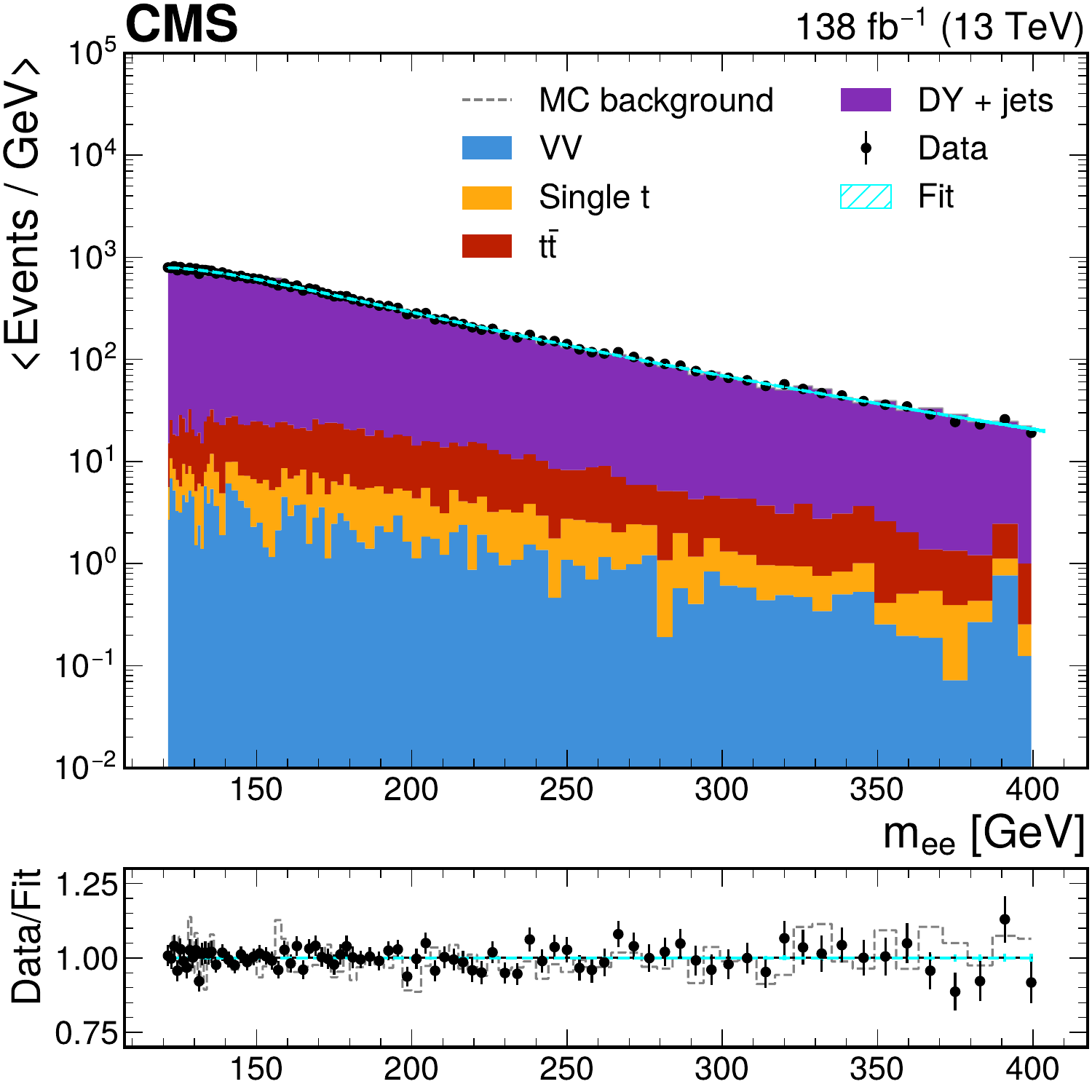}

  \caption{The 1-jet CRs with data fits used for the \SR\ background prediction. Upper left: \CRmmj. Upper right: \CReeb. Lower: \CReej. Events are divided by the bin width. We observe a shape difference between data and simulation in \CRmmj. The ratio of the nominal MC background values (dashed line) and data to the ABCD prediction is shown as a dashed line in the ratio plot. The MC background is included as a comparison, and is not used for predictions. \label{fig:CRswithData}} 
\end{figure}

\begin{figure}[hbt]
  \centering
  \includegraphics[width=0.48\textwidth]{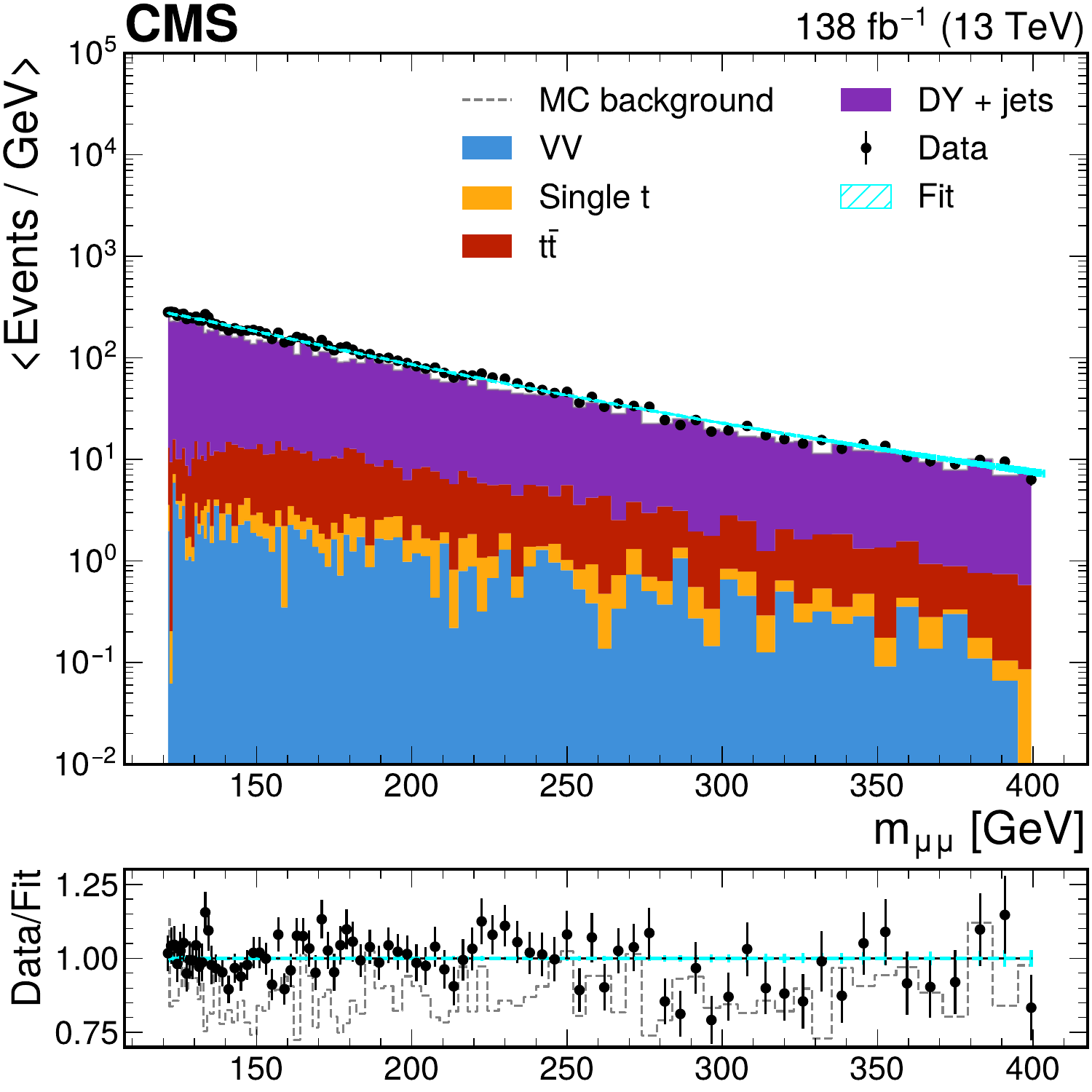}
  \hfill
  \includegraphics[width=0.48\textwidth]{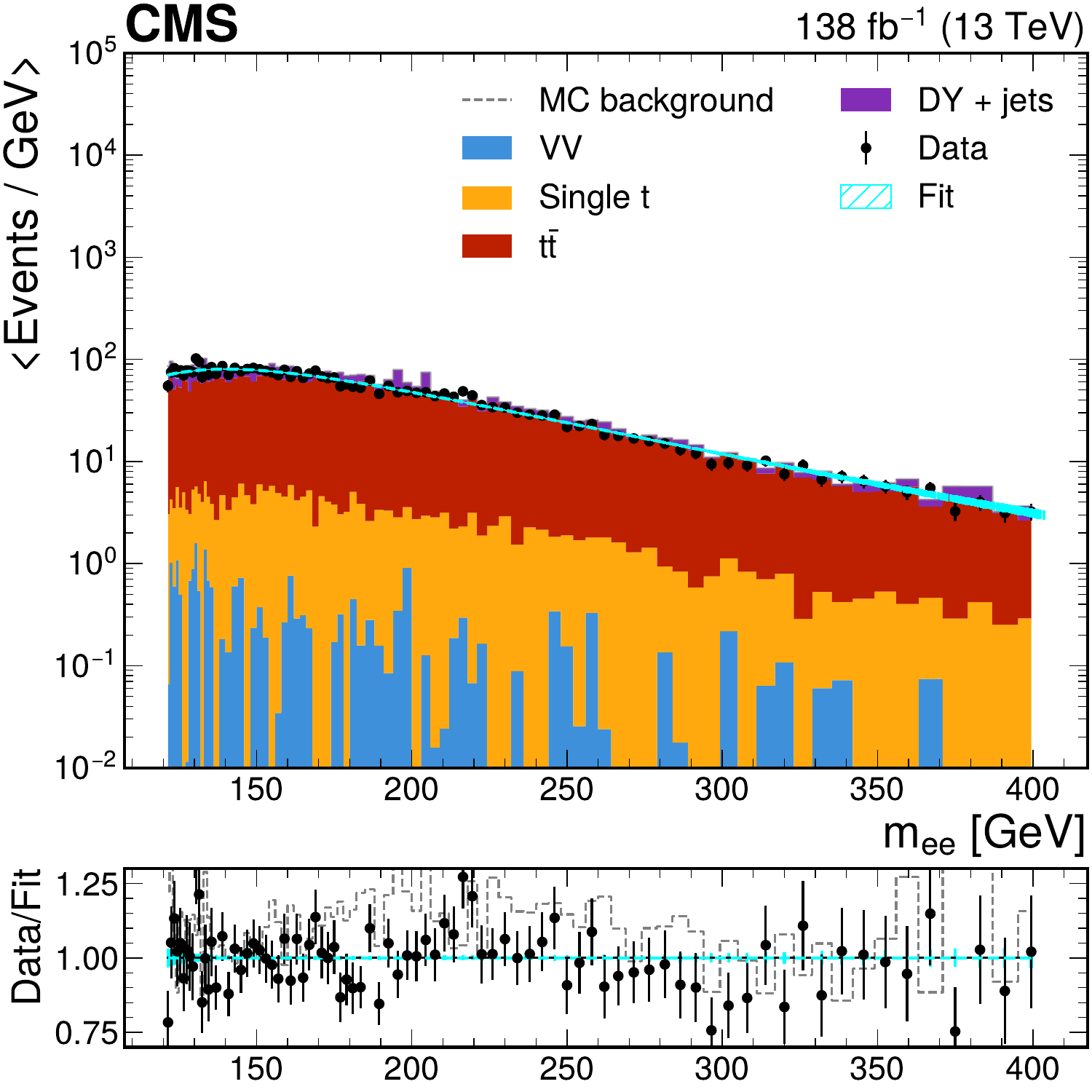}

  \vspace{1mm} 

  \includegraphics[width=0.48\textwidth]{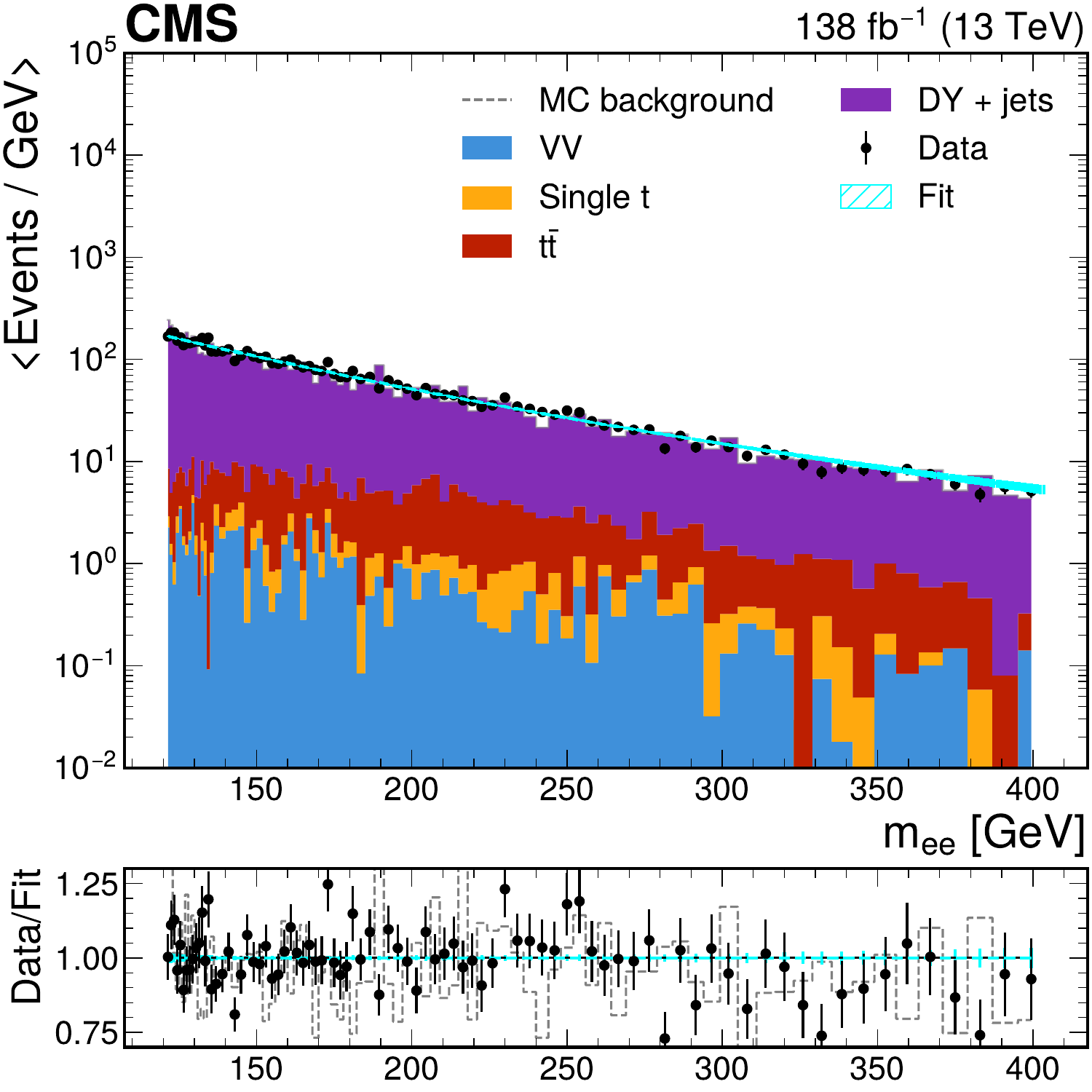}

  \caption{The 2-jet CRs with data fits used for the \SRTwo\ background prediction. Upper left: \CRmmjTwo. Upper right: \CReebTwo. Lower: \CReejTwo. Events are divided by the bin width. We observe a shape difference between the MC and data in \CRmmjTwo.  The ratio of the nominal MC background values (dashed line) and data to the ABCD prediction is shown as a dashed line in the ratio plot. The MC background is included as a comparison, and is not used for predictions.\label{fig:CRTwoswithData}}
\end{figure}

\begin{figure}[ht]
  \centering
    \includegraphics[width=0.48\textwidth]{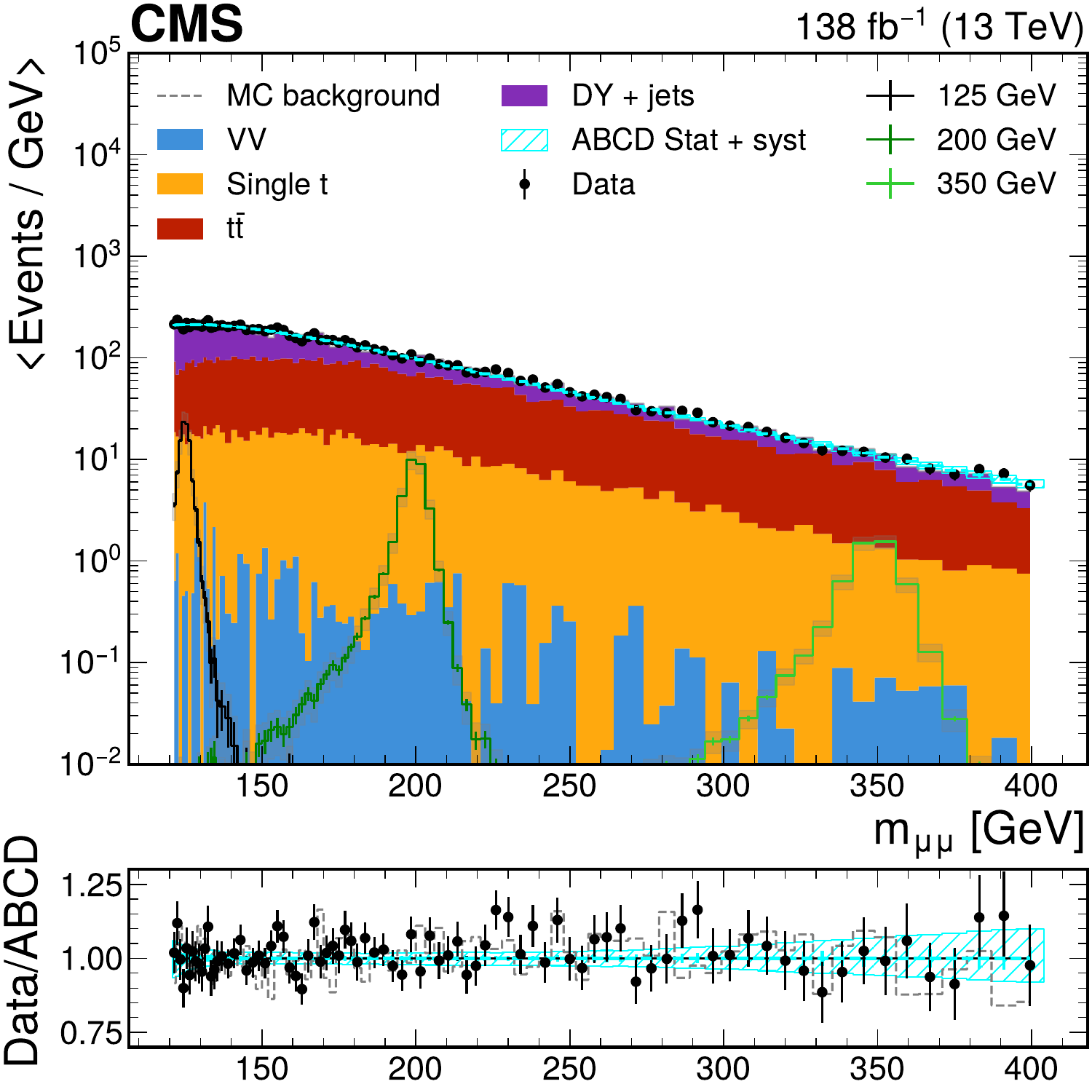}
    \hfill
    \includegraphics[width=0.48\textwidth]{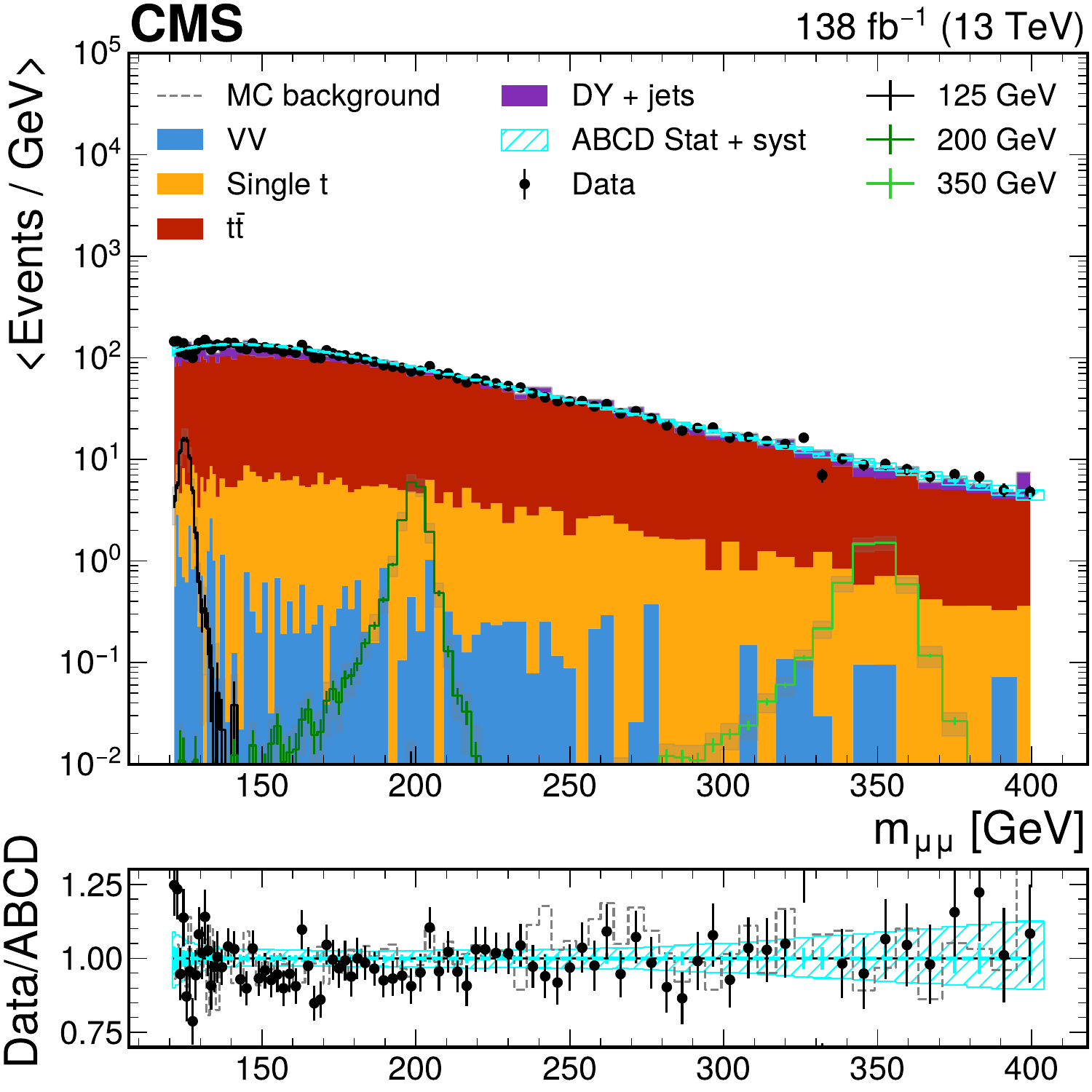}
   \caption{Distributions of \mll~in the \SR (left) and \SRTwo (right) SRs. Events are divided by the bin width. Simulated signal shapes for \PZpr boson masses of 125, 200, and 350\GeV are shown.  The ``Stat + syst" band shows the envelope of the fit variations with statistical uncertainties.  The ratio of the nominal MC background values (dashed line) and data to the ABCD prediction is shown as a dashed line in the ratio plot.  The MC background uncertainties are not shown for visual clarity.  \label{fig:DiLepMassFinal}}
\end{figure}

We utilize two types of \mll binning in this analysis.
We use a flat 1\GeV binning for the background estimation based on CRs in data to improve the fit stability in the CRs. For the SRs, we use a binning scheme that progressively widens in 1\GeV steps, starting from 1\GeV and ending at 9\GeV, to match the width of the \PZpr boson mass resonance peak. This ensures sufficient resolution for limit setting while avoiding low counts per bin. The dilepton mass distribution is plotted as the number of events normalized by the bin width (Figs. \ref{fig:BgrClosure}-\ref{fig:SR2}). The bin widths are rounded down to the closest integer of the reconstructed \PZpr boson width in\GeV for a resonance at the corresponding \mll. The \mll locations at which the binning widens are: 136, 182, 224, 264, 299, 335, 363, and 395\GeV.

All shapes and normalizations are extracted using a log-normal function:

\begin{equation}
f(m_{\ell\ell},\sigma,\mu, n) =  {\frac {n}{\mll\sigma {\sqrt {2\pi }}}}\ \exp \left[-{\frac {(\ln (m_{\ell\ell})-\mu )^{2}}{2\sigma ^{2}}}\right],
\label{eq:LogNorm}
\end{equation}

where $\sigma$ is the turn-on width, $\mu$ is the mean, and $n$ the normalization.

The comparisons of simulated backgrounds, including all systematic uncertainties, with the ABCD SR background predictions derived from fits to the simulated CRs are shown in Fig. \ref{fig:BgrClosure}. The respective ratios of the ABCD predictions derived on simulated CRs to the actual simulated prediction in each SR are also shown.

Discrepancies are observed between data and simulation in the CRs, as shown in Figs. \ref{fig:CRswithData} and \ref{fig:CRTwoswithData}, which motivated us not to use a simulated background prediction in favor of a background prediction based solely on data. Figure \ref{fig:DiLepMassFinal} presents the ABCD prediction compared to the data, illustrating the accuracy of the background estimation technique based on the CRs in data.
Between the ABCD background estimation and the observed background in data, we calculate a $\chi^2$ of 83.0, 67.1, 124.0 (42.7, 71.9, 77.4) over 84 bins for 2016, 2017, 2018 \SR (\SRTwo), respectively.

\section{Systematic uncertainties\label{sec:SysUnc}}

Systematic uncertainties of both theoretical and experimental origin that affect signal simulation are listed in Table \ref{tab:systematics}. Background uncertainties in this analysis are not subject to simulation uncertainties, but rather uncertainties in the closure of the ABCD method. These uncertainties affect both the signal and background acceptance, as well as the dilepton mass distribution, which is the primary search variable. As we aim to provide model-independent results, we factorize the results into signal acceptances for different final-states. Template shapes for the nominal dilepton mass resonance and each of its systematic variations are used to set limits on the shape prior to applying the acceptances for each final-state jet multiplicity. It is important to note that the addition of initial-state b jet production does not change the mass shape of the resonance.

The integrated luminosities for the 2016, 2017, and 2018 data-taking periods have 1.2--2.5\% individual uncertainties~\cite{CMS-LUM-17-003,CMS-PAS-LUM-17-004,CMS-PAS-LUM-18-002}, while the overall uncertainty for the 2016--2018 period is 1.6\%.

Mismodeling of pileup is evaluated by varying the total inelastic pp cross section used to predict the pileup distribution in the data by $\pm4.6\%$ \cite{CMS:2018mlc}. This affects the signal normalization in the range of 0.30--2.44\%, correlated across years.

Simulated jet four-momenta are varied according to the jet energy scale and resolution uncertainties, resulting in a systematic uncertainty in the transverse momentum of 2.4--4.4\% for the jet energy scale and 2.6--7.2\% for the jet energy resolution per jet. The misidentification rate of genuine jets in the pileup identification between data and simulation is also varied within its uncertainty, contributing 0.7--1.7\% uncertainty in the event yield. Finally, the uncertainty in the \pt- and $\abs{\eta}$-dependent \PQb tagging efficiency for \PQb, \PQc, and light-flavor quarks or gluons is evaluated by varying data-to-simulation scale factors applied to simulation by their uncertainties. These variations yield a change in normalization and shape of 0.6--1.6\%.

In June of 2018, the HCAL experienced a readout loss covering approximately a $\Delta\phi = 40^\circ$ arc in the region $1.3 < \abs{\eta} < 3.0$ \cite{CMS:2023gfb}. In MC simulations, jets in this region are assigned an additional 0.20--0.22\% HCAL jet energy scale uncertainty to account for uncertainties in the modeling of this issue.

The fit function used for the mass distribution in each CR is subject to correlations by which fluctuations in low-mass bins can inversely affect the fit predictions in the high-mass range, and vice versa. For example, an upward fluctuation in the lower part of the \CReeb\ mass range could cause the fitted function to underestimate the data in the higher part of the \CReeb\ mass range. These effects lead to nontrivial and varied changes in the fit prediction, depending on the bin where the fluctuation occurs, which are not adequately accounted for by taking the envelope of the covariance matrix. As such, the variations in the fit from fluctuations at each bin are introduced as a systematic uncertainty. Because \CReeb\ and \CReebTwo\ are the least populated regions, and therefore the most sensitive to statistical fluctuations in any bin, we use these CRs to estimate the correlations in the fit function across bins caused by such fluctuations. The central prediction of the fit is used as a baseline for the per-bin estimated yield. The Poissonian uncertainty in that estimated yield is taken as the bin's fluctuation value. For each bin, we vary the number of events up and down by its fluctuation value and refit the distribution for each variation. The resulting changes to the fit prediction based on the fit variations are assigned as shape uncertainties over the dilepton mass range.  This method preserves the correlations between the bins along the entire fit prediction curve. To demonstrate the size of the fit variations, the envelope of all these variations adds an effective uncertainty that we refer to as the background fit uncertainty of 2 to 16\% (2 to 17\%) in \SR (\SRTwo).

For the signal acceptances of different jet multiplicities, we take into account the PDF, factorization, and renormalization scale uncertainties of each. The PDF uncertainties are evaluated by reweighting the signal and background events using the 68\% envelopes of the NNPDF 3.0 or 3.1 set of eigenvector variations as per the recommendation of the PDF4LHC working group ~\cite{Ball_2022}. The PDF uncertainties result in a $<0.1\%$ uncertainty in the acceptance, correlated across the data-taking years. Factorization and renormalization scales, $\mu_{\textrm{F}}$ and $\mu_{\textrm{R}}$, impact within each matrix element diagram the presence of further associated soft jets. The associated soft jets mostly affect the acceptance of final states with no inherent outgoing bottom quark. Uncertainties in this are estimated for each final states's acceptance by halving or doubling each scale independently and choosing the largest deviations from nominal in each direction among all permutations as the total uncertainty, excluding the two extreme variations. The scale variations introduce a relative variation of up to 7\%, with a mean value of 2\%.
An overview of all uncertainties considered for the dilepton mass shape of the \PZpr boson resonance and the background and which part of the analysis they affect, as well as whether there are correlations across the data-taking years, are summarized in Table \ref{tab:systematics}.

\begin{table}[ht]
\centering
\topcaption{Overview of systematic uncertainty sources and their ranges. Systematic uncertainties that are treated as fully or partially correlated across the data-taking years are labeled with an asterisk.  \label{tab:systematics}} 
\begin{tabular}{lllcc}
  \hline
  \multirow{2}*{Uncertainty source} & \multirow{2}*{Type} & \multirow{2}*{Processes} & \multicolumn{2}{c}{Range} \\
   &               &               &     \SR &  \SRTwo \\
  \hline
  Trigger eff. &                    Per event &         Signal &        \multicolumn{2}{c}{2.2--4.7\%} \\ 
  Background fit &                  Per event &         Backgr. est. &  2--16\% & 2--17\%\\
  Integrated luminosity &     Per event &         Signal &        \multicolumn{2}{c}{1.2--2.5\%}\\
  Muon reco., ident., and iso. eff. &   Per muon &          Signal &        \multicolumn{2}{c}{0.2--0.6\%} \\
  Muon energy scale &               Per muon (\pt) &    Signal &        \multicolumn{2}{c}{0.1--0.2\%} \\
  Jet energy scale &                Per jet (\pt) &     Signal &        \multicolumn{2}{c}{2.4--4.4\%} \\
  Jet energy resolution &           Per jet (\pt)&      Signal &        \multicolumn{2}{c}{2.6--7.2\%} \\
  HCAL inefficiency*  &          Per jet (\pt) &     Signal &        \multicolumn{2}{c}{0.20--0.22\%} \\
  b tagging efficiency* &    Per jet &           Signal &        \multicolumn{2}{c}{0.6--1.6\%} \\
  Pileup*&                          Per event &         Signal &        \multicolumn{2}{c}{0.30--2.44\%}\\
  Pileup jet ident. &       Per jet &           Signal &        \multicolumn{2}{c}{0.7--1.7\%}\\
  PDF &                             Per event &         Signal &        $<$0.02\% & $<$0.2\%\\
  Factorization/renormalization scales &       Per event &         Signal &        $<$0.2\% & $<$0.4\%\\
  \hline
 \end{tabular}
\end{table}

\section{Results\label{sec:Results}}

The SR data agree with the background predictions within the uncertainties. 

We report limits on the product of inclusive cross section and branching fraction. Acceptance times efficiency for individual production processes, primarily based on the flavors and multiplicities (\PQb and \PQs) of the initial-state radiation (ISR), are listed in Tables \ref{tab:Acceptance_0b1}--\ref{tab:Acceptance_2s}, such that the limits can be reinterpreted for any model based on these processes. Contributions from other jet flavors (light-flavor and charm quarks) are not considered for the categorization, as they do not couple to the \PZpr boson in the model considered.

The \mll distributions corresponding to the reconstructed dilepton masses in the 1- and 2-jet categories, combined over all three data-taking years, are shown in Fig. \ref{fig:DiLepMassFinal}, along with a few simulated \PZpr boson signal masses. Limits are based on the background estimation from control samples in data.  The dominant background in the low-mass range of \SR\ is DY production, while in all other areas of the \mll distribution \ttbar dominates, for either SR. 
Single top quark background contributes at the 5 to 10\% level and rarer diboson backgrounds have sub-percent contributions.

\begin{figure}[ht]
    \centering
    \includegraphics[width=.32\textwidth]{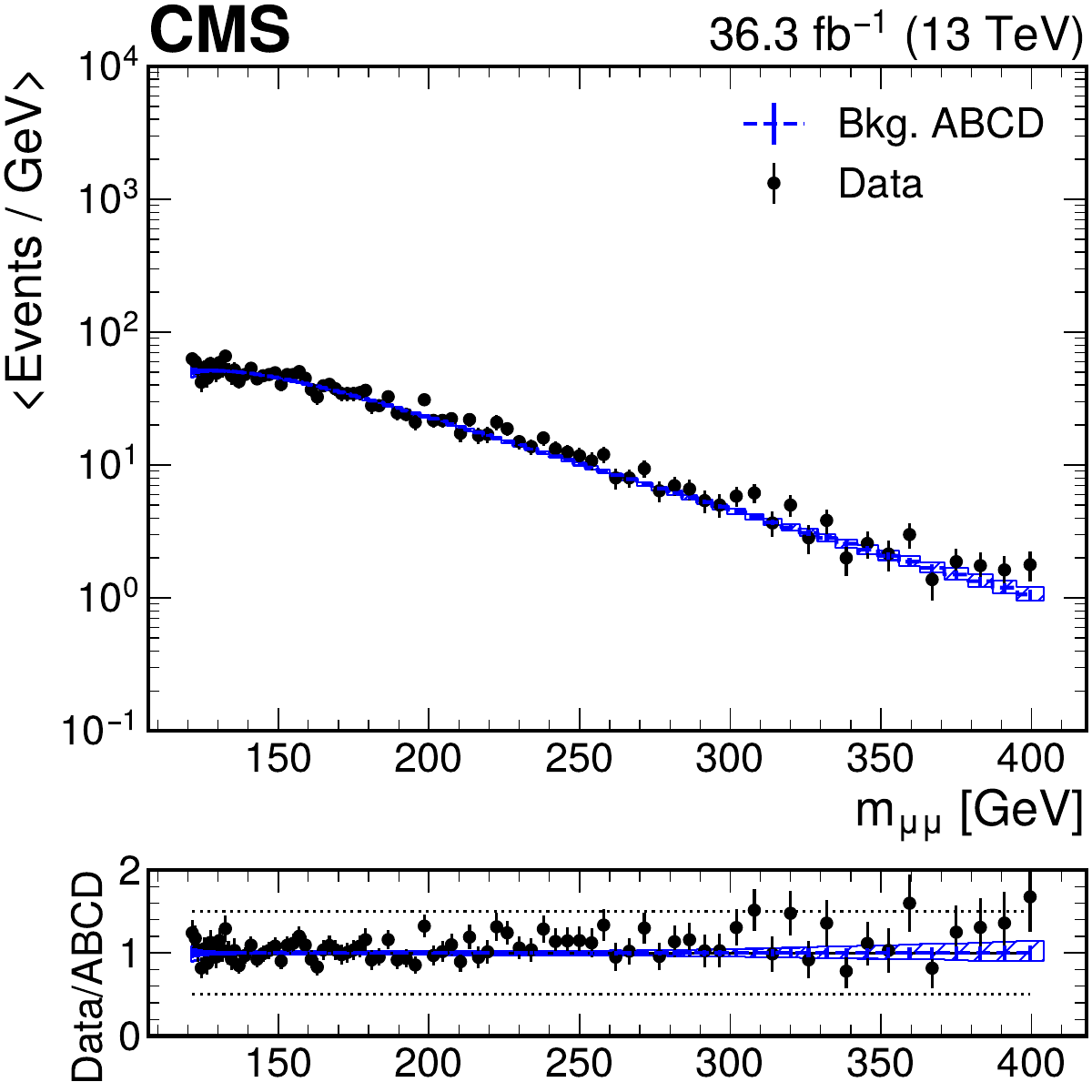}
    \hfill
    \includegraphics[width=.32\textwidth]{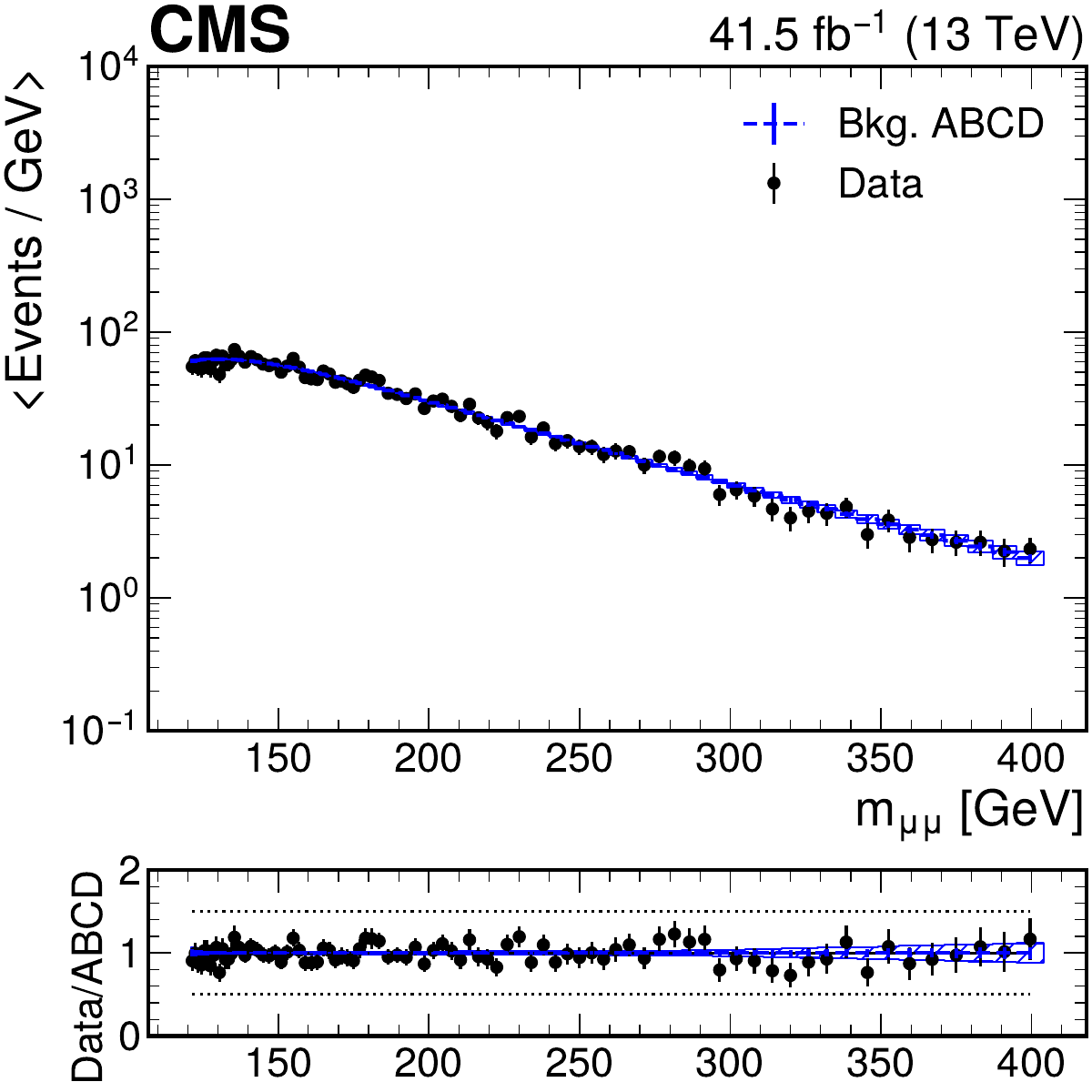}
    \hfill
    \includegraphics[width=.32\textwidth]{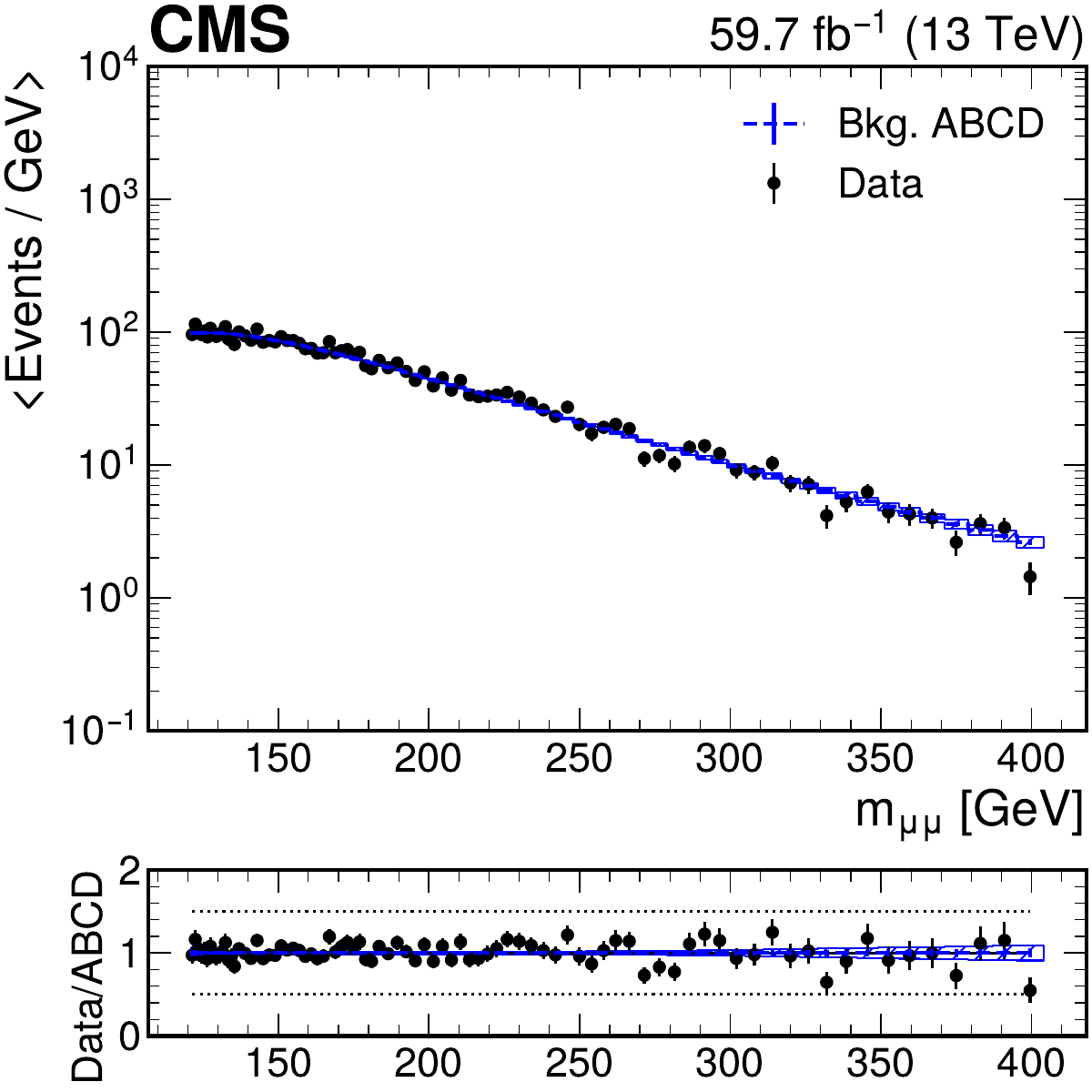}
     
     \caption{Data (black) vs. the ABCD method background prediction (blue) for 2016 (left), 2017 (middle), and 2018 (right) data-taking years in \SR. Events are divided by the bin width.  Error bars show statistical uncertainties of data. The blue band shows the propagated uncertainty of all individual fit variations in a given bin, which we consider to be uncorrelated. The lower panels show the ratio of the observed data to the background estimation.}
     \label{fig:SR1}
\end{figure}

\begin{figure}[ht]
    \centering

    \includegraphics[width=.32\textwidth]{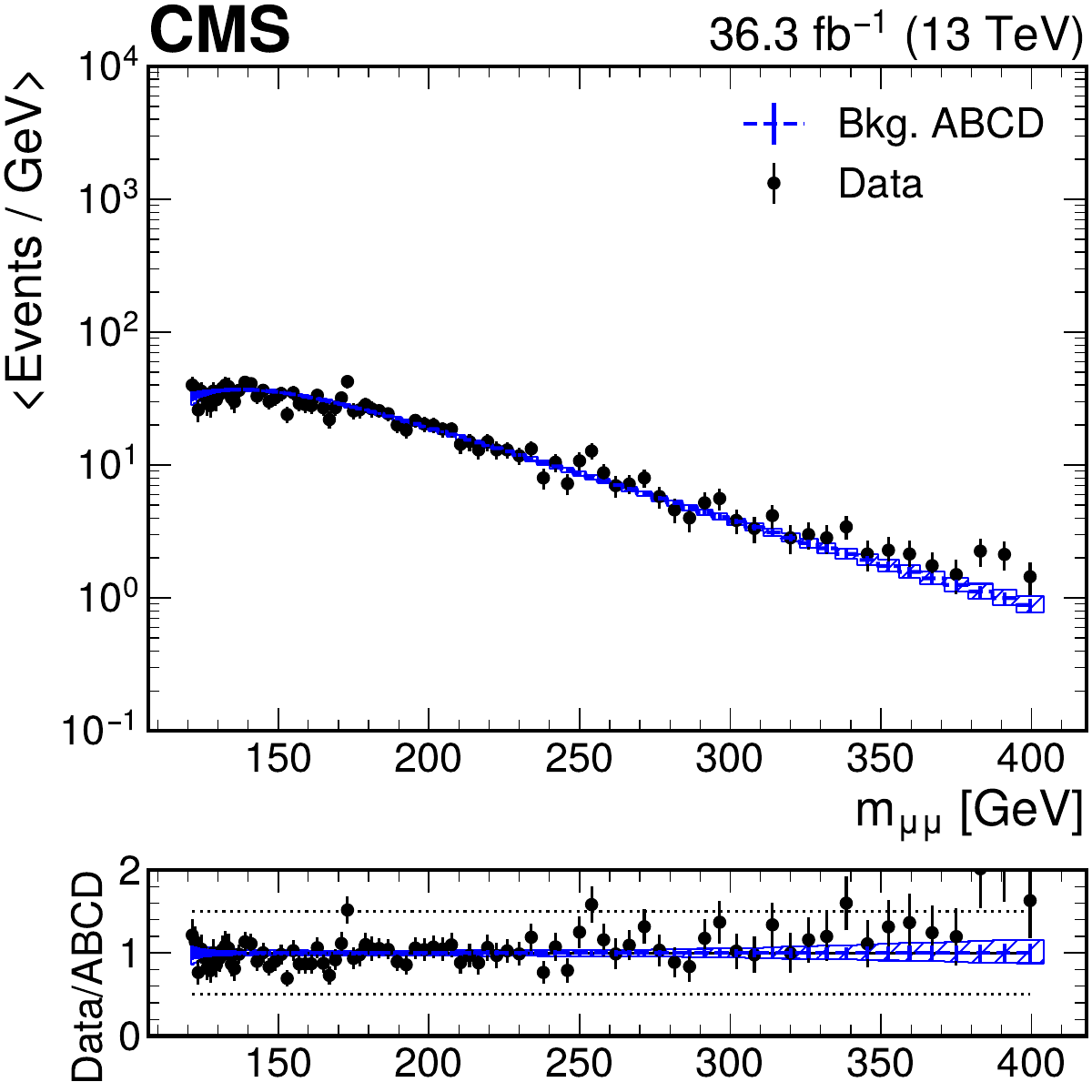}
    \hfill
    \includegraphics[width=.32\textwidth]{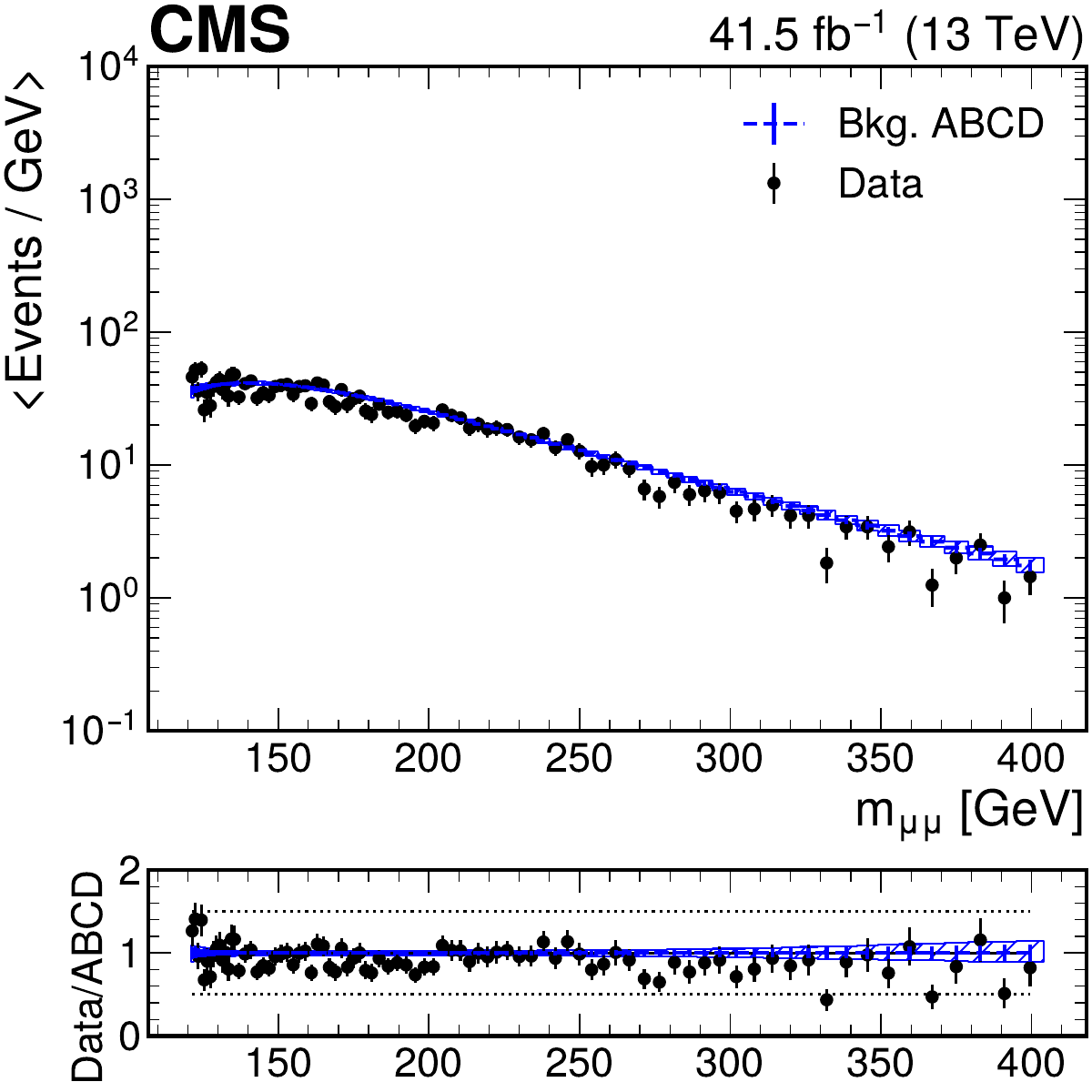}
    \hfill
    \includegraphics[width=.32\textwidth]{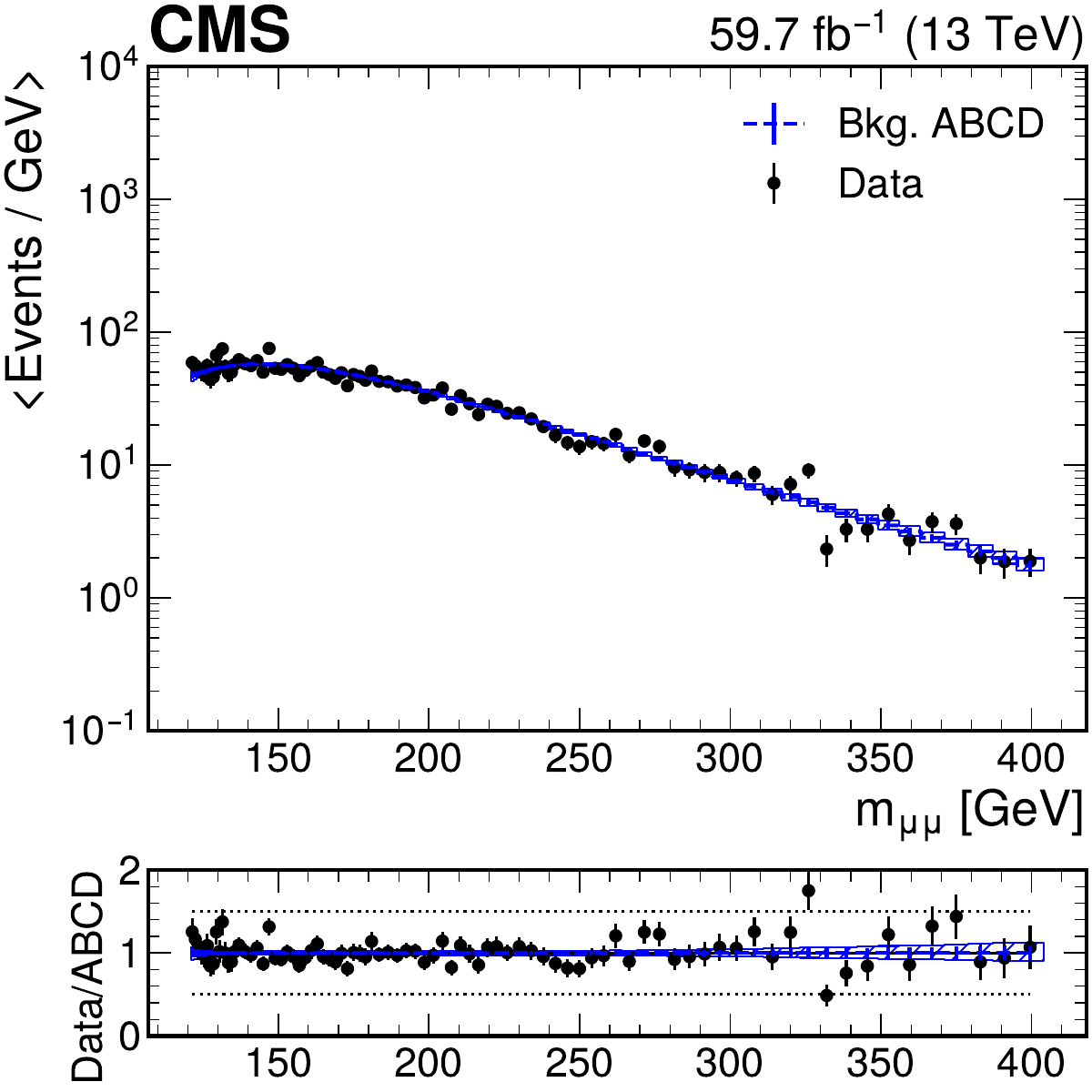}

     \caption{Data (black) vs. the ABCD method background prediction (blue) for 2016 (left), 2017 (middle), and 2018 (right) data-taking years in \SRTwo. Events are divided by the bin width. Error bars show statistical uncertainties of data. The blue band shows the propagated uncertainty of all individual fit variations in a given bin, which we consider to be uncorrelated. The lower panels show the ratio of the observed data to the background estimation. 
    }
     \label{fig:SR2}
\end{figure}

The total background shape and normalization have been estimated based on CRs in data, as described in Section \ref{sec:BackgroundEstimation}.  Asymptotic limits at 95\% confidence level (\CL) \cite{Cowan_2011} are set for \PZpr boson masses between 125 and 350\GeV, with corresponding reconstructed mass peak widths ranging from 1.7 to 7.2\GeV. The data compared to the ABCD method background predictions are shown in Figs. \ref{fig:SR1} and \ref{fig:SR2} for \SR\ and \SRTwo, and for each data taking year separately.

\begin{figure}[ht]

    \includegraphics[width=.5\linewidth]{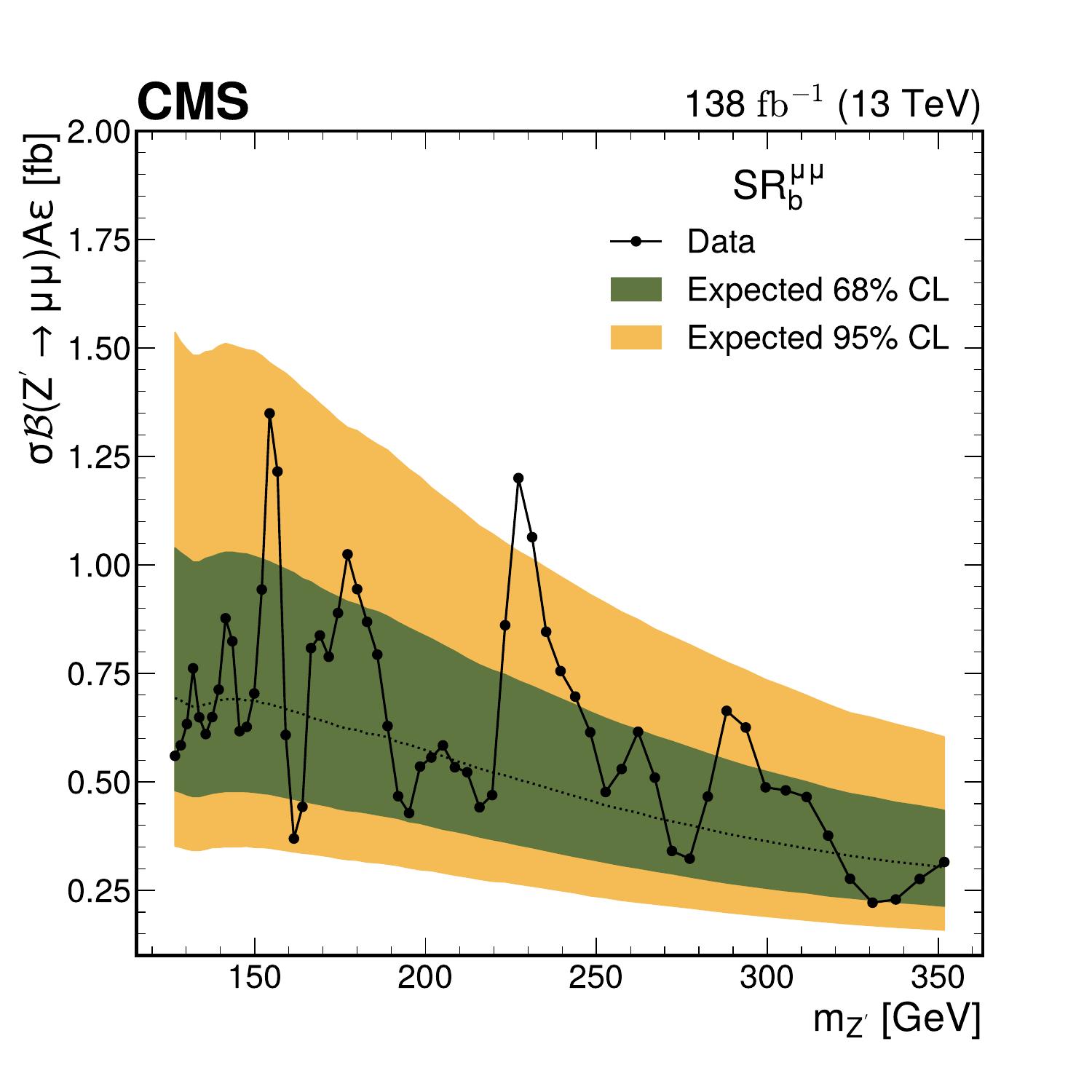}
    \hfill
    \includegraphics[width=.5\linewidth]{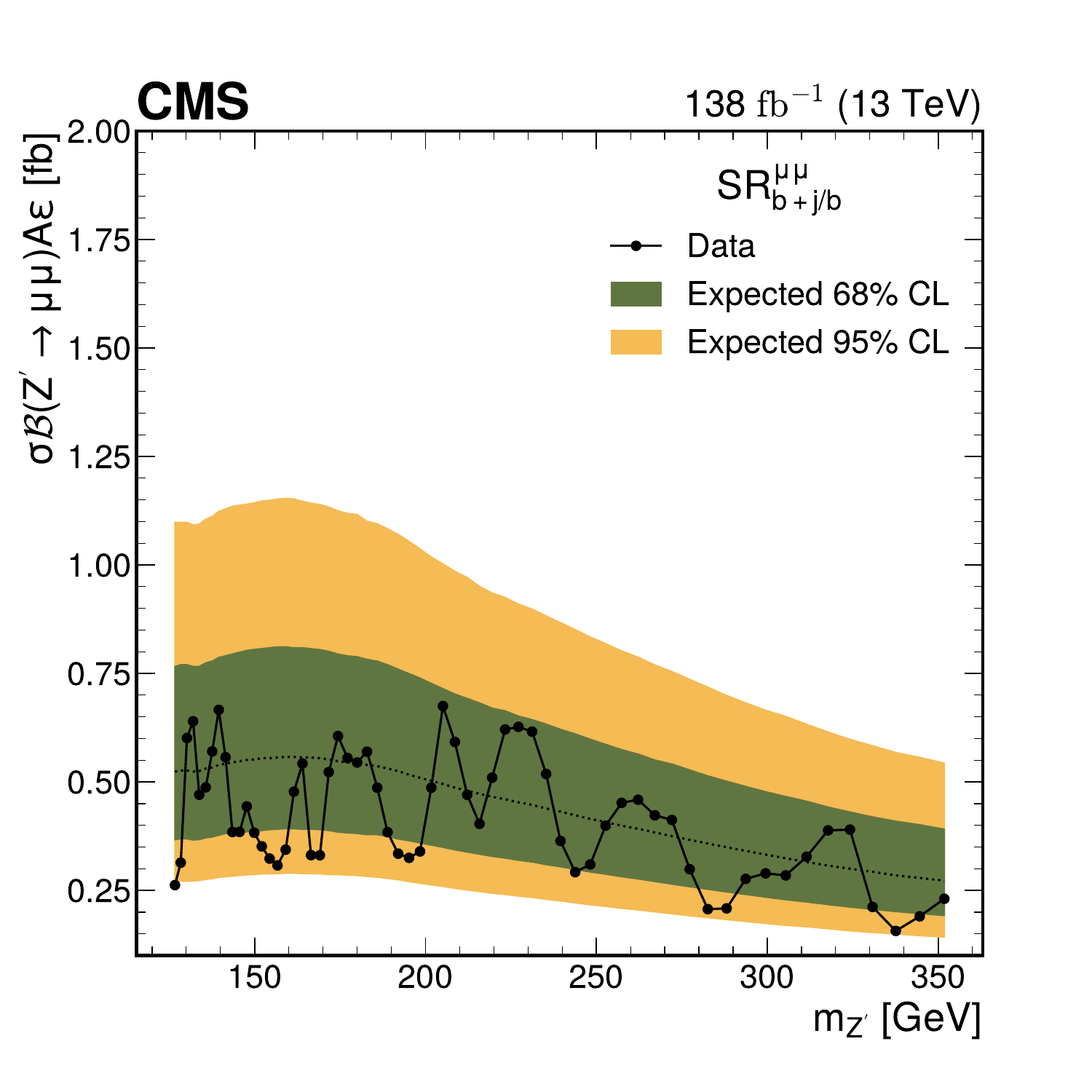}

 \caption{The asymptotic 95\% \CL limits on the product of cross section($\sigma$), branching fraction into a pair of muons ($\mathcal{B}$), acceptance (A), and efficiency ($\epsilon$). Left is \SR, right is \SRTwo. A combination of both limits depends on the relative acceptance contributions in each region and is omitted here. \label{fig:NewCountLimit}}
\end{figure}

A simultaneous maximum likelihood fit across data-taking years is performed in each jet multiplicity category separately to extract any potential signal contributions from the 0--2 parton signal processes shown in Fig. \ref{fig:Feynman}.
It has been shown that the dilepton mass shape is the same in our chosen benchmark model for all evaluated signal processes with the same \PZpr boson mass, regardless of the production diagram.
Systematic uncertainties as described in Section \ref{sec:SysUnc}~ are treated as nuisance parameters following log-normal probability density functions for purely rate-affecting cases, while systematic uncertainties affecting the \mll~ shape are assumed to follow a Gaussian probability density function instead. 
Asymptotic limits at 95\% \CL are set on the product of cross section, branching fraction, and acceptance for \PZpr boson masses between 125 and 350\GeV in steps equal to the reconstructed dimuon mass width. 
The limits are computed using the CMS statistical analysis tool \textsc{Combine}~\cite{CAT-23-001} which is based on the \textsc{RooFit}~\cite{Verkerke:2003ir} and \textsc{RooStats}~\cite{Moneta:2010pm} frameworks. 
The generator width for \PZpr boson masses of 125, 250, and 300\GeV is 0.6, 0.65, and 0.72\GeV, respectively.
Limits are also set on the presence of a signal for \SR\ and \SRTwo\ separately, as shown in Fig. \ref{fig:NewCountLimit}. No evidence of a statistically significant signal contribution is found from the fit to the data. The most notable deviation from the expected limits occurs in \SR\ at 230\GeV with a 3.1 standard deviations local significance; however, the global significance, computed using MC pseudo-experiments and accounting for both SRs, is only at 0.5 standard deviations due to the look-elsewhere effect \cite{Gross:2010qma}.
Since we aim to provide model-independent limits corresponding to different signal process diagrams, we do not combine the two SRs in this paper, to avoid assumptions on the mix of processes involved.

Production process categories we consider are based on matrix element level generator selections after hadronization and jet matching with the MLM prescription \cite{MLM_2007}, as described in Section \ref{sec:Samples}.
The primary concern of this categorization is the number of ISR jets and their flavor (bottom or strange), whilst a secondary concern is, for events with no hard matrix element ISR jets, the production of the \PZpr boson via one or two associated bottom quarks. The multiplicity of \PQb jets depends on the soft \PQb hadrons from hadronization in the underlying event, and therefore affects the acceptance.

Contributions from other jet flavors, gluons or light and charm quarks, are not considered for the categorization, as they do not couple to the \PZpr boson in the model considered.
The processes are grouped into the following categories, in the order shown in Fig. \ref{fig:Feynman}:
\begin{itemize}
    \item 0\PQb(\PQb-\PQs): No associated final-state \PQs or \PQb jet production, where the initial state involved a single bottom quark (first diagram of Fig. \ref{fig:Feynman}, tree level).
    \item 0\PQb(\PQb-\PQb): No associated final-state \PQs or \PQb jet production, where the initial state involved two different bottom quarks (first diagram of Fig. \ref{fig:Feynman}, tree level).
    \item 1\PQb: A single associated final-state \PQb jet (second diagram of Fig. \ref{fig:Feynman}).
    \item 1\PQs: A single associated final-state \PQs jet (second diagram of Fig. \ref{fig:Feynman}).
    \item 1\PQb+1\PQs: A pair of associated final-state \PQs and \PQb jets (third diagram of Fig. \ref{fig:Feynman}). 
    \item 2\PQb: Two associated final-state \PQb jets (third diagram of Fig. \ref{fig:Feynman}).
    \item 2\PQs: Two associated final-state \PQs jets (originating from an ISR gluon that splits into an $\PQs\overline{\PQs}$ pair). 
\end{itemize}

The acceptances times efficiency for each production category are given separately in Tables \ref{tab:Acceptance_0b1}--\ref{tab:Acceptance_2s} for each SR, including all relevant event selections. We note that the acceptances are not derived from a single b-tag efficiency; in addition to jets originating from b quarks in the hard process, additional b-tagged jets from gluon splitting or initial- and final-state radiation are also included, as well as effects from other object efficiencies and event selections. However, any realistic model interpretation will have to take into account not only the contribution of each diagram, its acceptance, and the uncertainty in the acceptance, but also the uncertainty in the relative contribution of each category among all categories with ISR and final-state radiation (FSR), and PDF uncertainties The ISR/FSR and PDF uncertainties can be large in comparison to the systematic uncertainties in the background estimation and the acceptance.

\begin{table}[hbt!]
\topcaption{Acceptance times efficiency by the SR for events with no generator-level matrix element ISR jets of strange or bottom flavor. Exactly one bottom quark contributes directly to the \PZpr boson production vertex. Statistical and systematic uncertainties for this 0\PQb(\PQb-\PQs) category are listed.\label{tab:Acceptance_0b1}}
\centering
\renewcommand{\arraystretch}{1.2}
\begin{tabular}{ccc}
\hline
$m_\PZpr$ [\GeVns{}] &  A$\epsilon$(\SR) [\%]   &  A$\epsilon$(\SRTwo) [\%]             \\
\hline
125  &  $1.09\pm0.01_{-0.01}^{+0.02}$ &  $0.12\pm0.00_{-0.01}^{+0.01}$ \\
150  &  $2.71\pm0.02_{-0.02}^{+0.03}$ &  $0.29\pm0.00_{-0.01}^{+0.01}$ \\
175  &  $3.80\pm0.02_{-0.03}^{+0.05}$ &  $0.46\pm0.01_{-0.01}^{+0.02}$ \\
200  &  $4.45\pm0.02_{-0.04}^{+0.06}$ &  $0.56\pm0.01_{-0.01}^{+0.02}$ \\
250  &  $4.73\pm0.28_{-0.06}^{+0.09}$ &  $0.76\pm0.11_{-0.02}^{+0.04}$ \\
300  &  $5.44\pm0.27_{-0.06}^{+0.08}$ &  $1.01\pm0.12_{-0.03}^{+0.04}$ \\
350  &  $5.44\pm0.02_{-0.04}^{+0.04}$ &  $0.98\pm0.01_{-0.01}^{+0.01}$ \\
\hline
\end{tabular}
\end{table}

\begin{table}[hbt!]
\topcaption{Acceptance times efficiency by the SR for events with no generator-level matrix element ISR jets of strange or bottom flavor. Exactly two bottom quarks contribute directly to the \PZpr boson production vertex. Statistical and systematic uncertainties for this 0\PQb(\PQb-\PQb) category are listed.\label{tab:Acceptance_0b2}}
\centering
\renewcommand{\arraystretch}{1.2}
\begin{tabular}{ccc}
\hline
$m_\PZpr$ [\GeVns{}] &  A$\epsilon$(\SR) [\%]   &  A$\epsilon$(\SRTwo) [\%]             \\
\hline
125  &  $1.87\pm0.01_{-0.03}^{+0.06}$ &  $0.23\pm0.00_{-0.01}^{+0.02}$ \\
150  &  $4.78\pm0.02_{-0.03}^{+0.06}$ &  $0.59\pm0.01_{-0.02}^{+0.04}$ \\
175  &  $6.71\pm0.03_{-0.03}^{+0.10}$ &  $0.89\pm0.01_{-0.02}^{+0.05}$ \\
200  &  $7.80\pm0.03_{-0.10}^{+0.11}$ &  $1.12\pm0.01_{-0.04}^{+0.05}$ \\
250  &  $9.09\pm0.03_{-0.10}^{+0.15}$ &  $1.48\pm0.01_{-0.04}^{+0.06}$ \\
300  &  $9.58\pm0.03_{-0.11}^{+0.15}$ &  $1.73\pm0.01_{-0.04}^{+0.06}$ \\
350  &  $9.63\pm0.03_{-0.08}^{+0.09}$ &  $1.86\pm0.01_{-0.03}^{+0.03}$ \\
\hline
\end{tabular}
\end{table}

\begin{table}[hbt!]
\topcaption{Acceptance times efficiency by the SR for events with exactly one generator-level matrix element ISR jets of bottom flavor. Statistical and systematic uncertainties for this 1\PQb category are listed.\label{tab:Acceptance_1b}}
\centering
\renewcommand{\arraystretch}{1.2}
\begin{tabular}{ccc}
\hline
$m_\PZpr$ [\GeVns{}] &  A$\epsilon$(\SR) [\%]   &  A$\epsilon$(\SRTwo) [\%]             \\
\hline
125  &   $6.34\pm0.06_{-0.09}^{+0.17}$ &  $2.59\pm0.04_{-0.12}^{+0.22}$ \\
150  &  $10.95\pm0.07_{-0.07}^{+0.14}$ &  $4.23\pm0.04_{-0.14}^{+0.24}$ \\
175  &  $15.78\pm0.08_{-0.09}^{+0.22}$ &  $5.95\pm0.05_{-0.13}^{+0.29}$ \\
200  &  $18.97\pm0.09_{-0.20}^{+0.25}$ &  $7.30\pm0.05_{-0.21}^{+0.29}$ \\
250  &  $21.42\pm0.11_{-0.24}^{+0.37}$ &  $9.50\pm0.07_{-0.28}^{+0.41}$ \\
300  &  $21.60\pm0.09_{-0.25}^{+0.34}$ &  $9.99\pm0.06_{-0.26}^{+0.35}$ \\
350  &  $21.41\pm0.07_{-0.17}^{+0.20}$ &  $9.38\pm0.04_{-0.13}^{+0.13}$ \\
\hline
\end{tabular}
\end{table}

\begin{table}[hbt!]
\topcaption{Acceptance times efficiency by the SR for events with exactly one generator-level matrix element ISR jets of strange flavor. Statistical and systematic uncertainties for this 1\PQs category are listed.\label{tab:Acceptance_1s}}
\centering
\renewcommand{\arraystretch}{1.2}
\begin{tabular}{ccc}
\hline
$m_\PZpr$ [\GeVns{}] &  A$\epsilon$(\SR) [\%]   &  A$\epsilon$(\SRTwo) [\%]             \\
\hline
125  &   $0.28\pm0.04_{-0.00}^{+0.01}$ &  $1.10\pm0.08_{-0.05}^{+0.08}$ \\
150  &   $0.57\pm0.04_{-0.00}^{+0.01}$ &  $1.69\pm0.07_{-0.05}^{+0.09}$ \\
175  &   $0.87\pm0.05_{-0.01}^{+0.01}$ &  $2.41\pm0.08_{-0.07}^{+0.11}$ \\
200  &   $0.94\pm0.05_{-0.01}^{+0.01}$ &  $2.87\pm0.09_{-0.08}^{+0.11}$ \\
250  &   $1.36\pm0.83_{-0.01}^{+0.00}$ &  $5.61\pm1.41_{-0.19}^{+0.30}$ \\
300  &   $1.47\pm0.53_{-0.01}^{+0.02}$ &  $3.52\pm0.82_{-0.08}^{+0.11}$ \\
350  &   $1.15\pm0.04_{-0.01}^{+0.01}$ &  $3.65\pm0.07_{-0.05}^{+0.04}$ \\
\hline
\end{tabular}
\end{table}

\begin{table}[hbt!]
\topcaption{Acceptance times efficiency by the SR for events with exactly two generator-level matrix element ISR jets, one of which has strange, the other bottom flavor. Statistical and systematic uncertainties for this 1\PQb+1\PQs category are listed.\label{tab:Acceptance_1bp1s}}
\centering
\renewcommand{\arraystretch}{1.2}
\begin{tabular}{ccc}
\hline
$m_\PZpr$ [\GeVns{}] &  A$\epsilon$(\SR) [\%]   &  A$\epsilon$(\SRTwo) [\%]             \\
\hline
125  &  $0.68\pm0.21_{-0.01}^{+0.02}$ &   $4.52\pm0.50_{-0.20}^{+0.32}$ \\
150  &  $1.09\pm0.17_{-0.01}^{+0.01}$ &   $7.67\pm0.44_{-0.24}^{+0.40}$ \\
175  &  $2.09\pm0.21_{-0.01}^{+0.02}$ &  $10.01\pm0.48_{-0.30}^{+0.46}$ \\
200  &  $2.57\pm0.26_{-0.02}^{+0.03}$ &  $10.23\pm0.47_{-0.26}^{+0.36}$ \\
250  &  $1.60\pm1.60_{-0.01}^{+0.01}$ &  $11.39\pm4.71_{-0.38}^{+0.62}$ \\
300  &  $3.46\pm2.04_{-0.04}^{+0.06}$ &  $12.45\pm3.82_{-0.31}^{+0.40}$ \\
350  &  $2.88\pm0.16_{-0.02}^{+0.02}$ &  $12.82\pm0.34_{-0.15}^{+0.13}$ \\
\hline
\end{tabular}
\end{table}

\begin{table}[hbt!]
\topcaption{Acceptance times efficiency by the SR for events with exactly two generator-level matrix element ISR jets of bottom flavor. Statistical and systematic uncertainties for this 2\PQb category are listed.\label{tab:Acceptance_2b}}
\centering
\renewcommand{\arraystretch}{1.2}
\begin{tabular}{ccc}
\hline
$m_\PZpr$ [\GeVns{}] &  A$\epsilon$(\SR) [\%]   &  A$\epsilon$(\SRTwo) [\%]             \\
\hline
125  &  $1.84\pm0.17_{-0.03}^{+0.06}$ &   $5.24\pm0.28_{-0.26}^{+0.48}$ \\
150  &  $2.74\pm0.19_{-0.02}^{+0.04}$ &   $8.48\pm0.34_{-0.30}^{+0.51}$ \\
175  &  $3.74\pm0.20_{-0.01}^{+0.04}$ &  $11.21\pm0.36_{-0.20}^{+0.59}$ \\
200  &  $4.40\pm0.19_{-0.05}^{+0.06}$ &  $12.47\pm0.32_{-0.39}^{+0.50}$ \\
250  &  $4.65\pm0.20_{-0.05}^{+0.07}$ &  $14.68\pm0.36_{-0.42}^{+0.60}$ \\
300  &  $5.27\pm0.16_{-0.05}^{+0.07}$ &  $14.87\pm0.28_{-0.35}^{+0.45}$ \\
350  &  $4.97\pm0.14_{-0.04}^{+0.04}$ &  $15.55\pm0.26_{-0.23}^{+0.24}$ \\
\hline
\end{tabular}
\end{table}

\begin{table}[hbtp!]
\topcaption{Acceptance times efficiency by the SR for events with exactly two generator-level matrix element ISR jets of strange flavor. Statistical and systematic uncertainties for this 2\PQs category are listed.\label{tab:Acceptance_2s}}
\centering
\renewcommand{\arraystretch}{1.2}
\begin{tabular}{ccc}
\hline
$m_\PZpr$ [\GeVns{}] &  A$\epsilon$(\SR) [\%]   &  A$\epsilon$(\SRTwo) [\%]             \\
\hline
125  &  $0.93\pm0.34_{-0.01}^{+0.02}$ &   $2.05\pm0.51_{-0.09}^{+0.17}$ \\
150  &  $1.55\pm0.42_{-0.00}^{+0.01}$ &   $5.09\pm0.83_{-0.17}^{+0.31}$ \\
175  &  $2.45\pm0.50_{-0.02}^{+0.02}$ &   $5.85\pm0.79_{-0.13}^{+0.29}$ \\
200  &  $2.74\pm0.48_{-0.01}^{+0.01}$ &   $7.96\pm0.87_{-0.22}^{+0.29}$ \\
250  &  $4.66\pm0.71_{-0.06}^{+0.11}$ &  $10.00\pm1.12_{-0.30}^{+0.43}$ \\
300  &  $5.00\pm0.66_{-0.05}^{+0.06}$ &   $9.92\pm0.94_{-0.25}^{+0.34}$ \\
350  &  $2.84\pm0.37_{-0.01}^{+0.01}$ &   $7.54\pm0.60_{-0.10}^{+0.10}$ \\
\hline
\end{tabular}
\end{table}

The 95\% exclusion limits are derived assuming that the \PZpr boson decays solely to muons. Since the \PZpr boson could also decay to electrons, contributions of this process to the dielectron CRs could modify the background estimation. To test this, we assume an equal contribution to the dimuon signal in size and in shape to the b-quark-rich dielectron CRs. Then, the ABCD background estimation is rerun as described in Section \ref{sec:BackgroundEstimation}, including the systematic calculations. Thus modified background predictions are used to recompute exclusion limits. At 200 and 350\GeV, these exclusion limits remain unchanged. However, at the 125\GeV mass point, the observed exclusion limit would degrade by 10\%.

\section{Summary\label{sec:Summary}}

A search for a new neutral \PZpr gauge boson with nonuniversal couplings, decaying into a pair of muons and produced with one or two associated jets, where at least one is \PQb tagged, has been presented.  The analysis is based on LHC proton-proton collision data at $\sqrt{s} = 13\TeV$, collected by the CMS experiment during 2016--2018, corresponding to an integrated luminosity of 138\fbinv. Exclusion bounds prior to this analysis predominantly focused on \PZpr boson production with universal couplings through light-quark fusion, leading to less stringent constraints on this signature. This analysis provides a complementary approach by targeting \PZpr boson production through bottom quark fusion, improving the sensitivity to this less-constrained region.

The presented results set the first specific limits on the product of cross section, branching fraction to dimuons, acceptance, and efficiency for \PQb{}\PZpr and \PQb{}j\PZpr production with \PZpr decays into a pair of muons. The data are consistent with the background-only hypothesis, showing no evidence of a signal.
Limits at 95\% confidence level on the product of cross section, branching fraction to dimuons, acceptance, and efficiency, ranging from 0.2 to 2\unit{fb}, are set for \PZpr boson masses between 125 and 350\GeV. 
Process-dependent products of acceptance and efficiency, and model-independent limits on the signal yield are provided.
These limits are presently the only results in the mass range of 125--200\GeV.
In the mass range of 200--350\GeV, these results provide the most stringent constraints on the specified set of bottom quark fusion production modes, serving as a complementary approach to inclusive \PZpr boson searches.

\begin{acknowledgments}
We congratulate our colleagues in the CERN accelerator departments for the excellent performance of the LHC and thank the technical and administrative staffs at CERN and at other CMS institutes for their contributions to the success of the CMS effort. In addition, we gratefully acknowledge the computing centers and personnel of the Worldwide LHC Computing Grid and other centers for delivering so effectively the computing infrastructure essential to our analyses. Finally, we acknowledge the enduring support for the construction and operation of the LHC, the CMS detector, and the supporting computing infrastructure provided by the following funding agencies: SC (Armenia), BMBWF and FWF (Austria); FNRS and FWO (Belgium); CNPq, CAPES, FAPERJ, FAPERGS, and FAPESP (Brazil); MES and BNSF (Bulgaria); CERN; CAS, MoST, and NSFC (China); MINCIENCIAS (Colombia); MSES and CSF (Croatia); RIF (Cyprus); SENESCYT (Ecuador); ERC PRG, TARISTU24-TK10 and MoER TK202 (Estonia); Academy of Finland, MEC, and HIP (Finland); CEA and CNRS/IN2P3 (France); SRNSF (Georgia); BMFTR, DFG, and HGF (Germany); GSRI (Greece); NKFIH (Hungary); DAE and DST (India); IPM (Iran); SFI (Ireland); INFN (Italy); MSIT and NRF (Republic of Korea); MES (Latvia); LMTLT (Lithuania); MOE and UM (Malaysia); BUAP, CINVESTAV, CONACYT, LNS, SEP, and UASLP-FAI (Mexico); MOS (Montenegro); MBIE (New Zealand); PAEC (Pakistan); MES, NSC, and NAWA (Poland); FCT (Portugal);  MESTD (Serbia); MICIU/AEI and PCTI (Spain); MOSTR (Sri Lanka); Swiss Funding Agencies (Switzerland); MST (Taipei); MHESI (Thailand); TUBITAK and TENMAK (T\"{u}rkiye); NASU (Ukraine); STFC (United Kingdom); DOE and NSF (USA).

\hyphenation{Rachada-pisek} Individuals have received support from the Marie-Curie program and the European Research Council and Horizon 2020 Grant, contract Nos.\ 675440, 724704, 752730, 758316, 765710, 824093, 101115353, 101002207, 101001205, and COST Action CA16108 (European Union); the Leventis Foundation; the Alfred P.\ Sloan Foundation; the Alexander von Humboldt Foundation; the Science Committee, project no. 22rl-037 (Armenia); the Fonds pour la Formation \`a la Recherche dans l'Industrie et dans l'Agriculture (FRIA) and Fonds voor Wetenschappelijk Onderzoek contract No. 1228724N (Belgium); the Beijing Municipal Science \& Technology Commission, No. Z191100007219010, the Fundamental Research Funds for the Central Universities, the Ministry of Science and Technology of China under Grant No. 2023YFA1605804, the Natural Science Foundation of China under Grant No. 12061141002, 12535004, and USTC Research Funds of the Double First-Class Initiative No.\ YD2030002017 (China); the Ministry of Education, Youth and Sports (MEYS) of the Czech Republic; the Shota Rustaveli National Science Foundation, grant FR-22-985 (Georgia); the Deutsche Forschungsgemeinschaft (DFG), among others, under Germany's Excellence Strategy -- EXC 2121 ``Quantum Universe" -- 390833306, and under project number 400140256 - GRK2497; the Hellenic Foundation for Research and Innovation (HFRI), Project Number 2288 (Greece); the Hungarian Academy of Sciences, the New National Excellence Program - \'UNKP, the NKFIH research grants K 131991, K 133046, K 138136, K 143460, K 143477, K 146913, K 146914, K 147048, 2020-2.2.1-ED-2021-00181, TKP2021-NKTA-64, and 2021-4.1.2-NEMZ\_KI-2024-00036 (Hungary); the Council of Science and Industrial Research, India; ICSC -- National Research Center for High Performance Computing, Big Data and Quantum Computing, FAIR -- Future Artificial Intelligence Research, and CUP I53D23001070006 (Mission 4 Component 1), funded by the NextGenerationEU program (Italy); the Latvian Council of Science; the Ministry of Education and Science, project no. 2022/WK/14, and the National Science Center, contracts Opus 2021/41/B/ST2/01369, 2021/43/B/ST2/01552, 2023/49/B/ST2/03273, and the NAWA contract BPN/PPO/2021/1/00011 (Poland); the Funda\c{c}\~ao para a Ci\^encia e a Tecnologia, grant CEECIND/01334/2018 (Portugal); the National Priorities Research Program by Qatar National Research Fund; MICIU/AEI/10.13039/501100011033, ERDF/EU, "European Union NextGenerationEU/PRTR", and Programa Severo Ochoa del Principado de Asturias (Spain); the Chulalongkorn Academic into Its 2nd Century Project Advancement Project, the National Science, Research and Innovation Fund program IND\_FF\_68\_369\_2300\_097, and the Program Management Unit for Human Resources \& Institutional Development, Research and Innovation, grant B39G680009 (Thailand); the Kavli Foundation; the Nvidia Corporation; the SuperMicro Corporation; the Welch Foundation, contract C-1845; and the Weston Havens Foundation (USA).
\end{acknowledgments}

\bibliography{auto_generated}
\cleardoublepage \appendix\section{The CMS Collaboration \label{app:collab}}\begin{sloppypar}\hyphenpenalty=5000\widowpenalty=500\clubpenalty=5000
\cmsinstitute{Yerevan Physics Institute, Yerevan, Armenia}
{\tolerance=6000
V.~Chekhovsky, A.~Hayrapetyan, V.~Makarenko\cmsorcid{0000-0002-8406-8605}, A.~Tumasyan\cmsAuthorMark{1}\cmsorcid{0009-0000-0684-6742}
\par}
\cmsinstitute{Institut f\"{u}r Hochenergiephysik, Vienna, Austria}
{\tolerance=6000
W.~Adam\cmsorcid{0000-0001-9099-4341}, J.W.~Andrejkovic, L.~Benato\cmsorcid{0000-0001-5135-7489}, T.~Bergauer\cmsorcid{0000-0002-5786-0293}, S.~Chatterjee\cmsorcid{0000-0003-2660-0349}, K.~Damanakis\cmsorcid{0000-0001-5389-2872}, M.~Dragicevic\cmsorcid{0000-0003-1967-6783}, P.S.~Hussain\cmsorcid{0000-0002-4825-5278}, M.~Jeitler\cmsAuthorMark{2}\cmsorcid{0000-0002-5141-9560}, N.~Krammer\cmsorcid{0000-0002-0548-0985}, A.~Li\cmsorcid{0000-0002-4547-116X}, D.~Liko\cmsorcid{0000-0002-3380-473X}, I.~Mikulec\cmsorcid{0000-0003-0385-2746}, J.~Schieck\cmsAuthorMark{2}\cmsorcid{0000-0002-1058-8093}, R.~Sch\"{o}fbeck\cmsAuthorMark{2}\cmsorcid{0000-0002-2332-8784}, D.~Schwarz\cmsorcid{0000-0002-3821-7331}, M.~Sonawane\cmsorcid{0000-0003-0510-7010}, W.~Waltenberger\cmsorcid{0000-0002-6215-7228}, C.-E.~Wulz\cmsAuthorMark{2}\cmsorcid{0000-0001-9226-5812}
% ...
\par}
\vskip\cmsinstskip
\dag:~Deceased\\
$^{1}$Also at Yerevan State University, Yerevan, Armenia\\
$^{2}$Also at TU Wien, Vienna, Austria\\
% ...
\end{sloppypar}
\end{document}